\documentclass[aps,pre,reprint,superscriptaddress,amsmath,amssymb,twocolumn,preprintnumbers]{revtex4-1}

\usepackage{graphics,graphicx}

\begin{document}
\preprint{Physical Review E}

\title{Logarithmic and nonlogarithmic scaling laws of two-point statistics in wall turbulence}

\author{Hideaki Mouri}
\affiliation{Meteorological Research Institute, Nagamine, Tsukuba 305-0052, Japan}
\affiliation{Graduate School of Science, Kobe University, Rokkodai, Kobe 657-8501, Japan}
\author{Takeshi Morinaga}
\affiliation{Meteorological Research Institute, Nagamine, Tsukuba 305-0052, Japan}
\author{Toshimasa Yagi}
\affiliation{Meteorological Research Institute, Nagamine, Tsukuba 305-0052, Japan}
\author{Kazuyasu Mori}
\affiliation{Meteorological Research Institute, Nagamine, Tsukuba 305-0052, Japan}



\begin{abstract}
Wall turbulence has a sublayer where one-point statistics, e.g., the mean velocity and the variances of some velocity fluctuations, vary logarithmically with the distance from the wall. This logarithmic scaling is found here for two-point statistics or specifically two-point cumulants of those fluctuations by means of experiments in a wind tunnel. As for corresponding statistics of the rate of the energy dissipation, the scaling is found to be not logarithmic. We reproduce these scaling laws with some mathematics and also with a model of energy-containing eddies that are attached to the wall.
\end{abstract}

\maketitle

\section{Introduction} \label{S1}

Within a sublayer of wall turbulence of an incompressible fluid, one-point statistics such as the mean velocity and the variances of some velocity fluctuations vary logarithmically with the distance from the wall. This logarithmic scaling is unusual, contrasting to power laws and exponential laws found in many other systems. Hence, in wall turbulence, we are to study scaling laws of two-point statistics.

The configuration is as follows. We take the $x$--$y$ plane at the wall. The $x$ direction is that of the mean stream. While $U(z)$ denotes the mean velocity at a distance $z$ from the wall, $u(z)$ and $w(z)$ denote velocity fluctuations in the streamwise and the wall-normal directions. Turbulence is homogeneous in the streamwise direction. Its thickness $\delta$ is a constant. The two points considered here are those separated by a streamwise distance $r$.

Asymptotically in the limit of high Reynolds number, there is a sublayer at $z/\delta \rightarrow 0$ such that the momentum flux $\rho \langle -uw(z) \rangle$ is constant at a value of $\rho u_{\tau}^2$ \cite{my71}. Here $\rho$ is the mass density, $u_{\tau}$ is the friction velocity, and $\langle \cdot \rangle$ denotes an average. Even in an actual case over a smooth or rough wall at a high but yet finite Reynolds number, this constant-flux sublayer is still a good approximation for a range of distances $z$.

Throughout the constant-flux sublayer, the friction velocity $u_{\tau}$ serves as a characteristic velocity. Since there is no constant in units of length, the mean velocity $U$ obeys a relation $\partial U/\partial z \varpropto u_{\tau}/z$ \cite{ll59, s48, my71}. Then,
\begin{subequations} \label{eq1}
\begin{equation} \label{eq1a}
\frac{U(z)}{u_{\tau}} = c_U + d_U \ln \! \left( \frac{\delta}{z} \right)  \!
\ \ \mbox{with} \ \
d_U = - \frac{1}{\kappa} .
\end{equation}
Here $c_U$ is an integration constant. The von K\'arm\'an constant $\kappa$ appears to be universal. Its estimate of $0.39 \pm 0.02$ is common among various configurations of wall turbulence, e.g., pipe flows, channel flows, and boundary layers \cite{mmhs13}.

The same scaling exists for the variance of streamwise velocity fluctuations $\langle u^2(z) \rangle$. According to the attached-eddy hypothesis of Townsend \cite{t76}, i.e., a model of a random superposition of energy-containing eddies that are attached to the wall,
\begin{equation} \label{eq1b}
\frac{\langle u^2(z) \rangle}{u_{\tau}^2} = c_{u^2} + d_{u^2} \! \ln \! \left( \frac{\delta}{z} \right) \!  .
\end{equation}
This law has been confirmed recently by means of laboratory experiments and field observations \cite{mmhs13, hvbs12}. As for its constants, while $c_{u^2} \simeq 1.4$--$1.8$ in pipe flows is distinct from $c_{u^2} \simeq 2.0$--$2.5$ in channel flows and boundary layers, $d_{u^2} \simeq 1.2$--$1.3$ is common among them \cite{mmhs13, hvbs13, vhs15, ofsbta17, smhffhs18, hvbs12}.

Other scaling laws are also known. An example is the local rate per unit mass of the energy dissipation $\varepsilon (z)$. By equating its average $\langle \varepsilon \rangle$ to the mean rate of the energy production $\langle -uw \rangle \partial U/\partial z = u_{\tau}^3/\kappa z$ at each distance $z$ in the constant-flux sublayer \cite{ll59},
\begin{equation} \label{eq1c}
\frac{\langle \varepsilon(z) \rangle}{u_{\tau}^3/z} = c_{\varepsilon} 
\ \ \mbox{with} \ \
c_{\varepsilon} = \frac{1}{\kappa} .
\end{equation}
\end{subequations}
This is in accordance with the attached-eddy hypothesis \cite{m17}, albeit possibly not exact with a discrepancy of $\pm 10$\% in the actual flow \cite{lm15}.

Being analogous to such one-point statistics, two-point statistics would exhibit some scaling laws. These are expected to offer much more information about the wall turbulence.

We are to use cumulants $\langle \alpha^n \beta^m \rangle_c$ of random variables $\alpha$ and $\beta$ at $n$ and $m = 0$, $1$, $2$, ... \cite{my71, ks77}. They are related to usual moments $\langle \alpha^l \beta^k \rangle$ at $l$ and $k = 0$, $1$, $2$, ....  For example \cite{c51},
\begin{subequations} \label{eq2}
\begin{align}
\label{eq2a} \langle \alpha          \rangle_c &=   \langle \alpha           \rangle                                                                  , \\
\label{eq2b} \langle \alpha  \beta   \rangle_c &=   \langle \alpha   \beta   \rangle - \langle \alpha         \rangle   \langle \beta \rangle         , \\
\label{eq2c} \langle \alpha^2\beta^2 \rangle_c &=   \langle \alpha^2 \beta^2 \rangle 
                                                 -  \langle \alpha^2         \rangle   \langle        \beta^2 \rangle 
                                                 -2 \langle \alpha   \beta   \rangle^2                                                        \nonumber \\
                                               & -2 \langle \alpha^2 \beta   \rangle   \langle \beta          \rangle   
                                                 -2 \langle \alpha           \rangle   \langle \alpha \beta^2 \rangle
                                                 +8 \langle \alpha           \rangle   \langle \alpha \beta   \rangle   \langle \beta \rangle \nonumber \\
                                               & -6 \langle \alpha           \rangle^2 \langle        \beta   \rangle^2 
                                                 +2 \langle \alpha^2         \rangle   \langle        \beta   \rangle^2 
                                                 +2 \langle \alpha           \rangle^2 \langle        \beta^2 \rangle. 
\end{align}
\end{subequations}
At $n+m \ge 2$, each moment $\langle \alpha^n \beta^m \rangle$ is contaminated nonlinearly with lower order moments. If all of such contamination is removed, the result is the cumulant $\langle \alpha^n \beta^m \rangle_c$. It is also identical to $\langle (\alpha - \langle \alpha \rangle )^n (\beta - \langle \beta \rangle)^m \rangle_c$. For a sum of independent random variables $\alpha_1$ and $\alpha_2$, each cumulant is identical to the sum of cumulants of the variables, i.e., $\langle (\alpha_1+\alpha_2)^n \beta^m \rangle_c = \langle \alpha_1^n \beta^m \rangle_c + \langle \alpha_2^n \beta^m \rangle_c$.

This linear character of cumulants would lead to simple scaling laws in the constant-flux sublayer.  Actually from the attached-eddy hypothesis \cite{m17},
\begin{subequations} \label{eq3}
\begin{equation} \label{eq3a}
\frac{\langle u^n(x,z)u^m(x+r,z) \rangle_c}{u_{\tau}^{n+m}} = C_{u^nu^m} \! \! \left( \frac{r}{z} \right) + D_{u^nu^m} \! \! \left( \frac{r}{z} \right) \! \ln \! \left( \frac{\delta}{z} \right)
\end{equation}
and
\begin{equation} \label{eq3b}
\frac{\langle \varepsilon^n(x,z) \varepsilon^m(x+r,z) \rangle_c}{(u_{\tau}^3/z)^{n+m}} = C_{\varepsilon^n \varepsilon^m} \! \! \left( \frac{r}{z} \right) \! .
\end{equation}
\end{subequations}
Here $x$ has become a dummy parameter because the turbulence is homogeneous in the streamwise direction. The functions $C_{u^nu^m}$, $D_{u^nu^m}$, and $C_{\varepsilon^n\varepsilon^m}$ are not yet determined but are related to constants of the one-point statistics. For example, $c_{u^2}$ in Eq.~(\ref{eq1b}) is identical to $C_{uu}(0)$ as well as to $C_{u^2}(0) = C_{u^2u^0}(0)$, while $c_{\varepsilon}$ in Eq.~(\ref{eq1c}) is identical to $C_{\varepsilon}(0) = C_{\varepsilon \varepsilon^0}(0)$.

We note that Eq.~(\ref{eq3a}) has been derived without adding any assumption to the original hypothesis of Townsend \cite{t76}. Although previous studies added assumptions \cite{pc82, dnk06, mizuno18, mm19}, they are not consistent with that hypothesis if their results do not satisfy Eq.~(\ref{eq3a}), aside from whether they appear reasonable or not \cite{m17,m19}.

The scaling laws of such two-point statistics are studied here. With some mathematics (Sec.~\ref{S2}), Eq.~(\ref{eq3}) is reproduced as an extension of Eq.~(\ref{eq1a}) for the mean velocity $U$. Then, by using data obtained from our experiments of boundary layers (Sec.~\ref{S3}), we confirm Eq.~(\ref{eq3}) in Sec.~\ref{S4}. Since its functions $C_{u^nu^m}$, $D_{u^nu^m}$, and $C_{\varepsilon^n\varepsilon^m}$ are not dependent on the above mathematics, they are discussed in terms of the attached-eddy hypothesis (Sec.~\ref{S5}). Finally, we conclude with remarks in Sec. \ref{S6}.

\section{Theory} \label{S2}

Our theory is to extend the scaling law of Eq.~(\ref{eq1a}) for the mean velocity $U$. Within the constant-flux sublayer at $z/\delta \rightarrow 0$, the local gradient $\partial U/ \partial z$ depends only on the local parameters $z$ and $u_{\tau}$ \cite{ll59, s48, my71}. Thus, we have used $\partial U/ \partial z \varpropto u_{\tau}/z$ to obtain Eq.~(\ref{eq1a}). Although  $\gamma \ne 1$ for $\partial U^{\gamma}/ \partial z \varpropto u_{\tau}^{\gamma}/z$ might appear equally plausible, this is not invariant under a Galilean transformation to add a constant to all of $U$ \cite{o01}.

The scaling law for any other statistics is required to be consistent with that for the mean velocity $U$. We rely on this requirement to constrain the former scaling via cumulants of the total streamwise velocity $U+u$. Another basis of our theory is that any nondimensional function is required to be described by nondimensional parameters alone.

With use of the friction velocity $u_{\tau}$, the total streamwise velocity $U+u$ is nondimensionalized as
\begin{equation} \label{eq4}
\alpha = \frac{U(z)+u(x,z)}{u_{\tau}}
\ \ \mbox{and} \ \
\beta = \frac{U(z)+u(x+r,z)}{u_{\tau}}.
\end{equation}
The distribution of $\alpha$ is described completely by its characteristic function $\langle e^{is\alpha} \rangle$ with a nondimensional parameter $s$ ranging from $-\infty$ to $+\infty$. By definition \cite{my71, ks77}, the cumulants $\langle \alpha^n \rangle_c$ are obtained from
\begin{subequations} \label{eq5}
\begin{equation} \label{eq5a}
\ln \langle e^{is\alpha} \rangle = \sum_{n=1}^{\infty} \langle \alpha^n \rangle_c \frac{(is)^n}{n!} 
\end{equation}
or from
\begin{equation} \label{eq5b}
 \langle \alpha^n \rangle_c = \left. \frac{\partial^n}{\partial (is)^n} \ln \langle e^{is\alpha} \rangle \right\vert_{s=0} .
\end{equation}
For consistency with $\partial U/ \partial z \varpropto u_{\tau}/z$, we impose
\begin{equation} \label{eq5c}
\frac{\partial}{\partial z} \ln \langle e^{is\alpha} \rangle = - \frac{\phi (s)}{z} .
\end{equation}
This relation is still invariant under the aforementioned Galilean transformation, which affects only a linear term of $\langle \alpha \rangle_c = \langle \alpha \rangle = U/u_{\tau}$ at $n=1$ in Eq.~(\ref{eq5a}). Since $\phi$ is a nondimensional function, it does not depend on $z$ that is not nondimensional. From Eqs.~(\ref{eq5b}) and (\ref{eq5c}),
\begin{equation} \label{eq5d}
\frac{\partial}{\partial z} \langle \alpha^n \rangle_c = - \left. \frac{1}{z} \frac{\partial^n \phi (s)}{\partial (is)^n} \right\vert_{s=0} = - \frac{\phi^{(n)}(0)}{i^n z}.
\end{equation}
\end{subequations}
We replace $\phi^{(n)}(0)/i^n$ with a constant $d_{(U+u)^n}$. Because of $\langle \alpha^n \rangle_c = \langle (U+u)^n \rangle_c/u_{\tau}^n$,
\begin{equation} \label{eq6}
\frac{\partial}{\partial z} \frac{\langle [U(z)+u(z)]^n \rangle_c}{u_{\tau}^n} = - \frac{d_{(U+u)^n}}{z}.
\end{equation}
The integration of Eq.~(\ref{eq6}) leads to the logarithmic laws of Eq.~(\ref{eq1a}) for $\langle U+u \rangle_c = U$ via $d_{U+u} = d_U = -1/\kappa$ and of Eq.~(\ref{eq1b}) for $\langle (U+u)^2 \rangle_c = \langle u^2 \rangle_c  = \langle u^2 \rangle$ via $d_{(U+u)^2} = d_{u^2}$. Such a law is obtained also for $\langle (U+u)^4 \rangle_c =\langle u^4 \rangle_c = \langle u^4 \rangle - 3 \langle u^2 \rangle^2$ and so on. However, the corresponding law for a moment $\langle u^l \rangle$ is usually not simple because $\langle u^l \rangle$ is contaminated nonlinearly with cumulants $\langle u^n \rangle_c$ of orders $n < l$ as inferred from Eq.~(\ref{eq2}).

To extend our theory into two-point cumulants, we use $\alpha$ and $\beta$ from Eq.~(\ref{eq4}). For their joint distribution \cite{my71, ks77}, the characteristic function is $\langle e^{is \alpha + it \beta} \rangle$. The cumulants $\langle \alpha^n \beta^m \rangle_c$ are obtained from
\begin{subequations} \label{eq7}
\begin{equation} \label{eq7a}
\ln \langle e^{is \alpha + it \beta} \rangle = \! \! \! \! \sum_{n+m=1}^{\infty} \! \! \! \langle \alpha^n \beta^m \rangle_c \frac{(is)^n}{n!} \frac{(it)^m}{m!}
\end{equation}
or from
\begin{equation} \label{eq7b}
 \langle \alpha^n \beta^m \rangle_c = \left. \frac{\partial^{n+m}}{\partial (is)^n \partial (it)^m} \ln \langle e^{is \alpha + it \beta} \rangle \right\vert_{s=t=0} .
\end{equation}
For consistency with Eq.~(\ref{eq5c}), we incorporate the separation of the two points $r$ as
\begin{equation} \label{eq7c}
\frac{\partial}{\partial z} \ln \langle e^{is \alpha + it \beta} \rangle = - \frac{\varphi (s,t,r/z)}{z} .
\end{equation}
Here $\varphi$ is a nondimensional function of the nondimensional parameters $s$, $t$, and $r/z$. They are independent of one another and also of the other parameter $z$. From Eqs.~(\ref{eq7b}) and (\ref{eq7c}),
\begin{align} \label{eq7d}
\frac{\partial}{\partial z} \langle \alpha^n \beta^m \rangle_c & = \left. - \frac{1}{z} \frac{\partial^{n+m}\varphi (s,t,r/z)}{\partial (is)^n\partial (it)^m } \right\vert_{s=t = 0} \nonumber \\
                                                                                             & =         - \frac{\varphi^{(n,m,0)} (0,0,r/z)}{i^{n+m} z}.
\end{align}
\end{subequations}
We consider the cases of $n+m \ge 2$ and use the function $D_{u^nu^m}(r/z)$ in place of $\varphi^{(n,m,0)} (0,0,r/z)/i^{n+m}$. Because of $\langle \alpha^n \beta^m \rangle_c = \langle u^n(x,z)u^m(x+r,z) \rangle_c/u_{\tau}^{n+m}$,
\begin{equation} \label{eq8}
\frac{\partial}{\partial z} \frac{\langle u^n(x,z)u^m(x+r,z) \rangle_c}{u_{\tau}^{n+m}} = - \frac{D_{u^nu^m}(r/z)}{z} .
\end{equation}
The integration of Eq.~(\ref{eq8}) leads to the logarithmic law of Eq.~(\ref{eq3a}).

For an extension into any other quantity, a joint distribution between $\alpha$ for this quantity and $\beta$ for $(U+u)/u_{\tau}$ is imposed to satisfy Eq.~(\ref{eq7}). Since Eq.~(\ref{eq7}) is reduced to Eq.~(\ref{eq5}) if $m=0$, the quantity itself satisfies Eq.~(\ref{eq5}) and then Eq.~(\ref{eq7}). We thereby obtain its own scaling laws.

The laws for fluctuations of the spanwise velocity are logarithmic as in the case of the streamwise velocity $u$. For the wall-normal velocity $w$, since it is equal to $0$ at the wall \cite{t76}, we adopt $\phi \equiv \varphi \equiv 0$ in Eqs.~(\ref{eq5}) and (\ref{eq7}). Hence, the one-point cumulants are constants. The two-point cumulants are functions of $r/z$ alone. Both of them are independent of $\ln (\delta /z)$. The same laws have been derived from the attached-eddy hypothesis \cite{t76,m17}.

We also consider the local rate of the energy dissipation $\varepsilon$. Although the dissipation is due to the fluid viscosity $\nu$, its value does not affect statistics of $\varepsilon$ such as the average $\langle \varepsilon \rangle$ if the Reynolds number is high enough \cite{ll59, t35, igk09}. The rate  $\varepsilon$ is nondimensionalized as
\begin{equation} \label{eq9}
\alpha = \frac{\varepsilon (x,z)}{u_{\tau}^3/z} \ \ \mbox{and} \ \ \beta = \frac{\varepsilon (x+r,z)}{u_{\tau}^3/z}.
\end{equation}
The mean rate of the energy production $u_{\tau}^3/\kappa z$ is enhanced in the limit $z/\delta \rightarrow 0$ of the constant-flux sublayer. With respect to this, the dissipation rate $\varepsilon$ needs to remain finite. We accordingly adopt $\phi \equiv \varphi \equiv 0$ in Eqs.~(\ref{eq5}) and (\ref{eq7}). Via integration, Eq.~(\ref{eq5d}) at $n=1$ yields Eq. (\ref{eq1c}) if the integration constant $c_{\varepsilon}$ is assigned to be $1/\kappa$, while Eq.~(\ref{eq7d}) yields Eq.~(\ref{eq3b}).

Thus, without invoking the attached eddies, we have reproduced the functional forms of Eqs.~(\ref{eq1}) and (\ref{eq3}). Such a scaling law exists simply because the constant-flux sublayer has no parameter except for the distance $z$ and the friction velocity $u_{\tau}$. The laws are limited to cumulants. Only their laws are extended systematically from that for the mean velocity $U$. If allowed by the condition at the wall surface, the law becomes logarithmic. For the existence of these scaling laws, an attached eddy is unnecessary, albeit useful to discussing their functions $C_{u^nu^m}$, $D_{u^nu^m}$, and $C_{\varepsilon^n \varepsilon^m}$ (see Sec.~\ref{S5}).

\begin{figure*}
\begin{center}
\rotatebox{90}{
\resizebox{5.50cm}{!}{\includegraphics*[9.9cm,1.9cm][18.5cm,29.2cm]{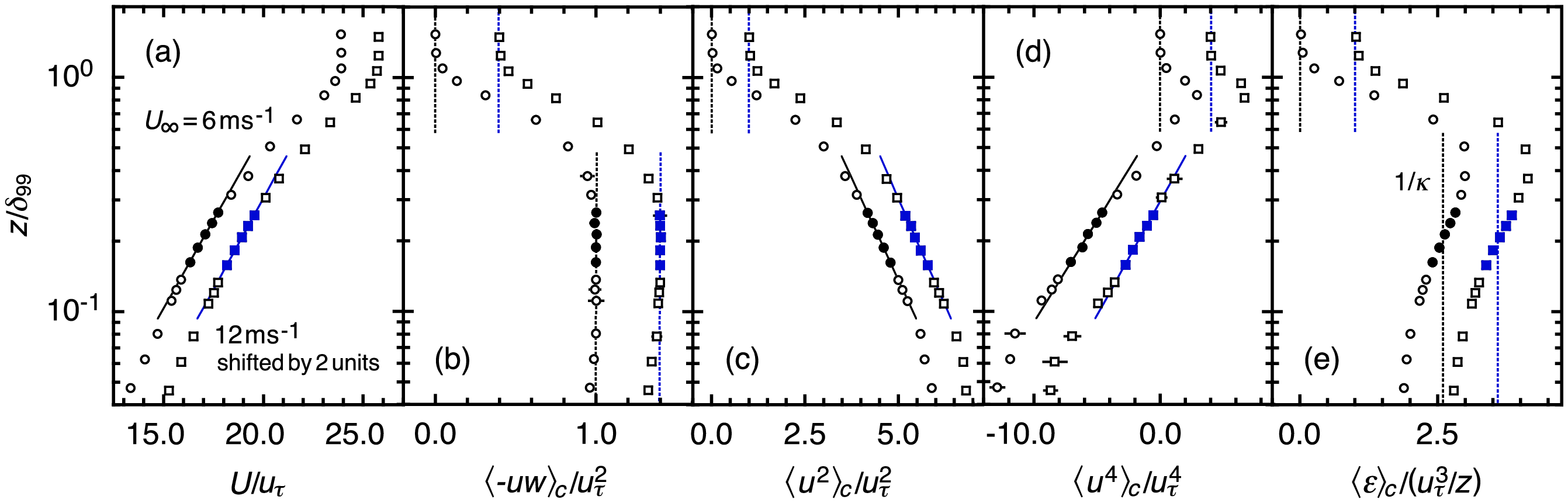}}
}
\caption{\label{f1} One-point statistics (a) $U(z)/u_{\tau}$, (b) $\langle -uw(z) \rangle_c/u_{\tau}^2 = \langle -uw(z) \rangle/u_{\tau}^2$, (c) $\langle u^2(z) \rangle_c/u_{\tau}^2 = \langle u^2(z) \rangle/u_{\tau}^2$, (d) $\langle u^4(z) \rangle_c/u_{\tau}^4 = [ \langle u^4(z) \rangle - 3 \langle u^2(z) \rangle^2]/u_{\tau}^4$, and (e) $\langle \varepsilon(z) \rangle_c/(u_{\tau}^3/z) = \langle \varepsilon(z) \rangle/(u_{\tau}^3/z)$ against $z/\delta_{99}$ for $U_{\infty} = 6$\,m\,s$^{-1}$ (circles) and $12$\,m\,s$^{-1}$ (squares shifted horizontally by two units). The filled symbols lie in the constant-flux sublayer. To these, solid lines are regression fits of Eq.~(\ref{eq1a}), (\ref{eq1b}), or (\ref{eq3a}). We provide $\pm 2 \sigma$ errors, albeit not including those for $u_{\tau}$ and $z-z_{\rm wt}$.}
\end{center}
\end{figure*} 

\section{Experiments} \label{S3}

Experiments of turbulent boundary layers were done in a wind tunnel of the Meteorological Research Institute. We use coordinates $x_{\rm wt}$, $y_{\rm wt}$, and $z_{\rm wt}$ in the streamwise, spanwise, and floor-normal directions. Their origin $x_{\rm wt} = y_{\rm wt} = z_{\rm wt} = 0$ is on the center of the floor at the upstream end of the test section of the tunnel. Its size is ${\mit\Delta}x_{\rm wt} = 18$\,m, ${\mit\Delta}y_{\rm wt} = 3$\,m, and ${\mit\Delta}z_{\rm wt} = 2$\,m. The cross section ${\mit\Delta}y_{\rm wt} \times {\mit\Delta}z_{\rm wt}$ is the same upstream to $x_{\rm wt} = -4$\,m.

Upon the entire floor from $x_{\rm wt} = -4$\,m to $+18$\,m with an interval of ${\mit\Delta}x_{\rm wt} = 0.1$\,m, spanwise rods of diameter $3.0$\,mm were set as roughness. It displaces the zero plane of the wall turbulence $z=0$ from the floor surface $z_{\rm wt}=0$ \cite{my71}, for which we assume $z - z_{\rm wt} = -1.5 \pm 1.5$\,mm \cite{losda03}.

The incoming flow velocity $U_{\infty}$ was set at $6$ or $12$\,m\,s$^{-1}$. Over some range of distances $z_{\rm wt}$ at $x_{\rm wt} = +14$ m and at $y_{\rm wt} = 0$\,m, where the turbulence had been well developed and had become almost independent of the position $x_{\rm wt}$, we measured the streamwise velocity $U+u$.

We used a hot-wire anemometer made up of a constant-temperature system (Dantec, 90C10) and of a single-wire probe (Dantec, 55P04). The wire was of platinum-plated tungsten, $5$\,$\mu$m in diameter, $1.25$\,mm in sensing length, and oriented to the spanwise direction. Its resistance overheat ratio was set at $0.80$. We calibrated the anemometer before and after each series of the measurements.

The anemometer signal was low-pass filtered and then digitally sampled. For each pair of $U_{\infty}$ and $z_{\rm wt}$, the sampling frequency $f_s$ was set as high as possible, provided that noise was still negligible at around $f_s$ where the energy spectrum of the signal had decayed substantially \cite{po97, mhk07}. The filter cutoff was at $f_s/2$.

The total length of the data at each of the distances $z_{\rm wt}$ from $45$ to $105$\,mm was as large as $3.2 \times 10^8$ for $U_{\infty} = 6$ m\,s$^{-1}$ under $f_s = 16$\,kHz and $4.0 \times 10^8$ for $U_{\infty} = 12$\,m\,s$^{-1}$ under $f_s = 44$\,kHz.  At the other distances $z_{\rm wt}$, we individually obtained $4.0 \times 10^7$ data.

During these measurements, we monitored the flow conditions such as the temperature. They are used to estimate the fluid viscosity $\nu$.

From our data along time $t_{\rm wt}$, spatial information is obtained via Taylor's hypothesis of $x = -Ut_{\rm wt}$. Despite some known problems \cite{dn08, dj09}, we rely on this hypothesis up to a large separation of $r/z \gtrsim 10^2$. The reason is the value of $\langle u^2 \rangle /U^2$. It was small enough, i.e., $\lesssim 0.03$, at all the distances $z_{\rm wt}$.

The local rate $\varepsilon$ of the energy dissipation is obtained as $15\nu (\partial_x u )^2$ by assuming local isotropy of the turbulence. For the calculation of the derivative $\partial_x u$, we use the four-point finite difference. This is a surrogate of the true rate, but its result for $\langle \varepsilon (x, z) \varepsilon (x+r, z) \rangle_c$ is reliable except at smallest separations $r$ around and below the Kolmogorov length $\eta = (\nu^3 / \langle \varepsilon \rangle)^{1/4}$ \cite{cgs03}.

These calculations are made for individual segments of length $10^7$ of our data. Among segments, statistics exhibit scatters. They originate in variations of experimental conditions, calibration uncertainties, and incomplete convergence due to a limited sampling time. After removing segments that are too noisy for some uncertain reason, we use these scatters to estimate the final errors in a standard manner \cite{br03}.

Supplementary short measurements were also done. To estimate the boundary layer thickness $\delta_{99}$, i.e., a distance $z$ at which $U$ is $99$\% of its maximum, we measured $U$ with an interval of ${\mit\Delta}z_{\rm wt} = 10\,\mbox{mm} \simeq 0.03\delta_{99}$. Furthermore, to obtain the momentum flux $\rho \langle -uw \rangle$, we measured $u$ and $w$ by utilizing a crossed-wire probe of the anemometer (Dantec, 55P53). Its wires were $1$\,mm in separation and oriented at $\pm 45^{\circ}$ to the streamwise direction. The other settings were the same as for the single-wire probe. We estimate the errors as described above about the long measurements.

\section{Results} \label{S4}

Figure \ref{f1} shows one-point statistics semilogarithmically as a function of $z/\delta = z/\delta_{99}$. Their parameters are summarized in Table \ref{t1}. For these and other following results, $\pm 2 \sigma$ errors are given as typical uncertainties \cite{br03}.

The constant-flux sublayer is observed in Fig.~\ref{f1} at least from $z/\delta_{99} \simeq 0.14$ to $0.28$ (filled symbols). Throughout this range, $U$ in Fig.~\ref{f1}(a) is logarithmic, $\langle -uw \rangle_c =  \langle -uw \rangle$ in Fig.~\ref{f1}(b) is constant, and $\langle u^2 \rangle_c = \langle u^2 \rangle$ in Fig.~\ref{f1}(c) is logarithmic.

\begin{table}[tbp]
\begingroup
\squeezetable
\caption{\label{t1} Parameters for $U_{\infty} = 6$ and $12$\,m\,s$^{-1}$: boundary layer thickness $\delta_{99}$, friction velocity $u_{\tau}$, viscosity $\nu$, Reynolds number $\delta_{99} u_{\tau}/\nu$, aerodynamic roughness $z_0$ and von K\'arm\'an constant $\kappa$ of $U(z)/u_{\tau} = \ln (z/z_0)/\kappa$ as well as $C_{u^n}(0)$ and $D_{u^n}(0)$ of Eq.~(\ref{eq3a}). The uncertainties are $\pm 2 \sigma$ errors. We also provide ranges of the Kolmogorov length $\eta$ and of the total sampling time $T$ among distances $z$ in the constant-flux sublayer.}
\begin{ruledtabular}
\begin{tabular}{lccc}
                             & Unit             & $U_{\infty} = 6$\,m\,s$^{-1}$ & $U_{\infty} = 12$\,m\,s$^{-1}$ \\  \hline
$\delta_{99}$                & mm               & $392   \pm  5$                & $403   \pm 5$                  \\    
$u_{\tau} $                  & mm\,s$^{-1}$     & $259   \pm  1$                & $512   \pm 0$                  \\
$\nu$                        & mm$^2$\,s$^{-1}$ & $15.0  \pm  0.0$              & $14.7  \pm 0.0$                \\
$\delta_{99} u_{\tau}/\nu$   & $10^3$           & $6.74  \pm  0.08$             & $14.0  \pm 0.2$                \\      
$\eta$                       & mm               & $0.27$ to  $0.29$             & $0.16$ to $0.17$               \\
$z_0$                        & mm               & $0.22  \pm  0.07$             & $0.19  \pm 0.07$               \\ 
$\kappa$                     &                  & $0.35  \pm  0.02$             & $0.36  \pm 0.02$               \\
$C_{u^2}(0)=C_{uu}(0)$       &                  & $2.52  \pm  0.15$             & $2.59  \pm 0.16$               \\
$D_{u^2}(0)=D_{uu}(0)$       &                  & $1.25  \pm  0.07$             & $1.19  \pm 0.08$               \\
$C_{u^4}(0)=C_{u^2u^2}(0)$   &                  & $2.08  \pm  1.18$             & $1.48  \pm 1.14$               \\
$D_{u^4}(0)=D_{u^2u^2}(0)$   &                  & $-5.01 \pm  0.48$             & $-4.47 \pm 0.46$               \\
$T U_{\infty} / \delta_{99}$ & $10^6$           & $0.28$ to  $0.31$             & $0.20$ to $0.24$               \\
\end{tabular}
\end{ruledtabular}
\endgroup
\end{table}
\begin{figure}[htbp]
\resizebox{7.84cm}{!}{\includegraphics*[3.0cm,1.7cm][17.5cm,28.4cm]{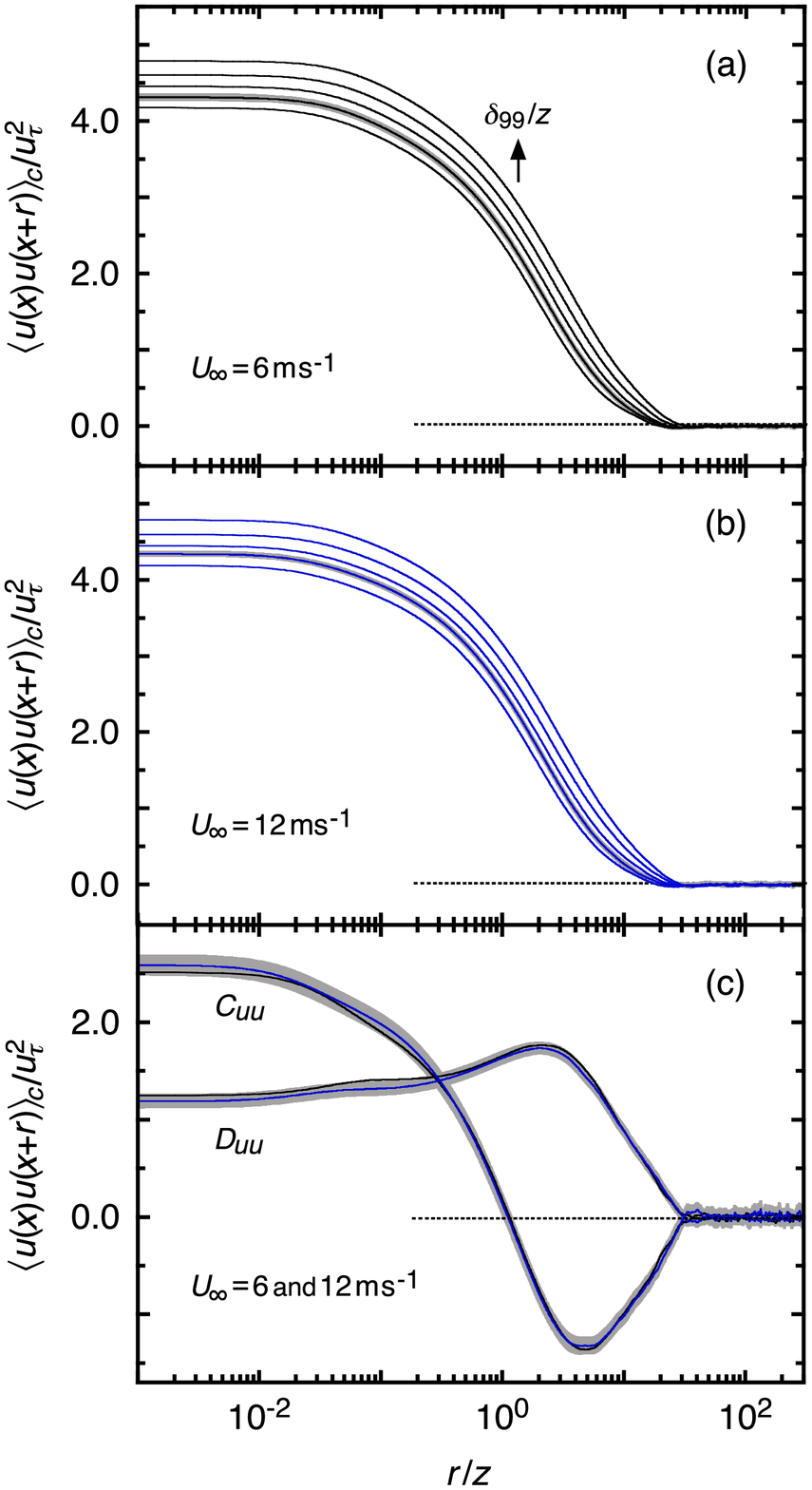}}
\caption{\label{f2} Two-point velocity cumulant in the constant-flux sublayer $\langle u(x,z) u(x+r,z) \rangle_c/u_{\tau}^2 = \langle u(x,z) u(x+r,z) \rangle/u_{\tau}^2$ against $r/z$ for (a) $U_{\infty} = 6$\,m\,s$^{-1}$ and (b) $12$\,m\,s$^{-1}$. The solid arrow indicates an increase in $\delta_{99}/z$. The gray areas are examples of $\pm 2 \sigma$ errors, albeit not including those for $u_{\tau}$ and $z-z_{\rm wt}$. The panel (c) shows $C_{uu}(r/z)$ and $D_{uu}(r/z)$ of Eq. (\ref{eq3a}). For $U_{\infty} = 12$\,m\,s$^{-1}$, we provide $\pm 2 \sigma$ errors including those for $u_{\tau}$ but not for $z-z_{\rm wt}$. }
\end{figure} 
\begin{figure}[bp]
\rotatebox{90}{
\resizebox{5.50cm}{!}{\includegraphics*[3.0cm,7.2cm][16.7cm,26.5cm]{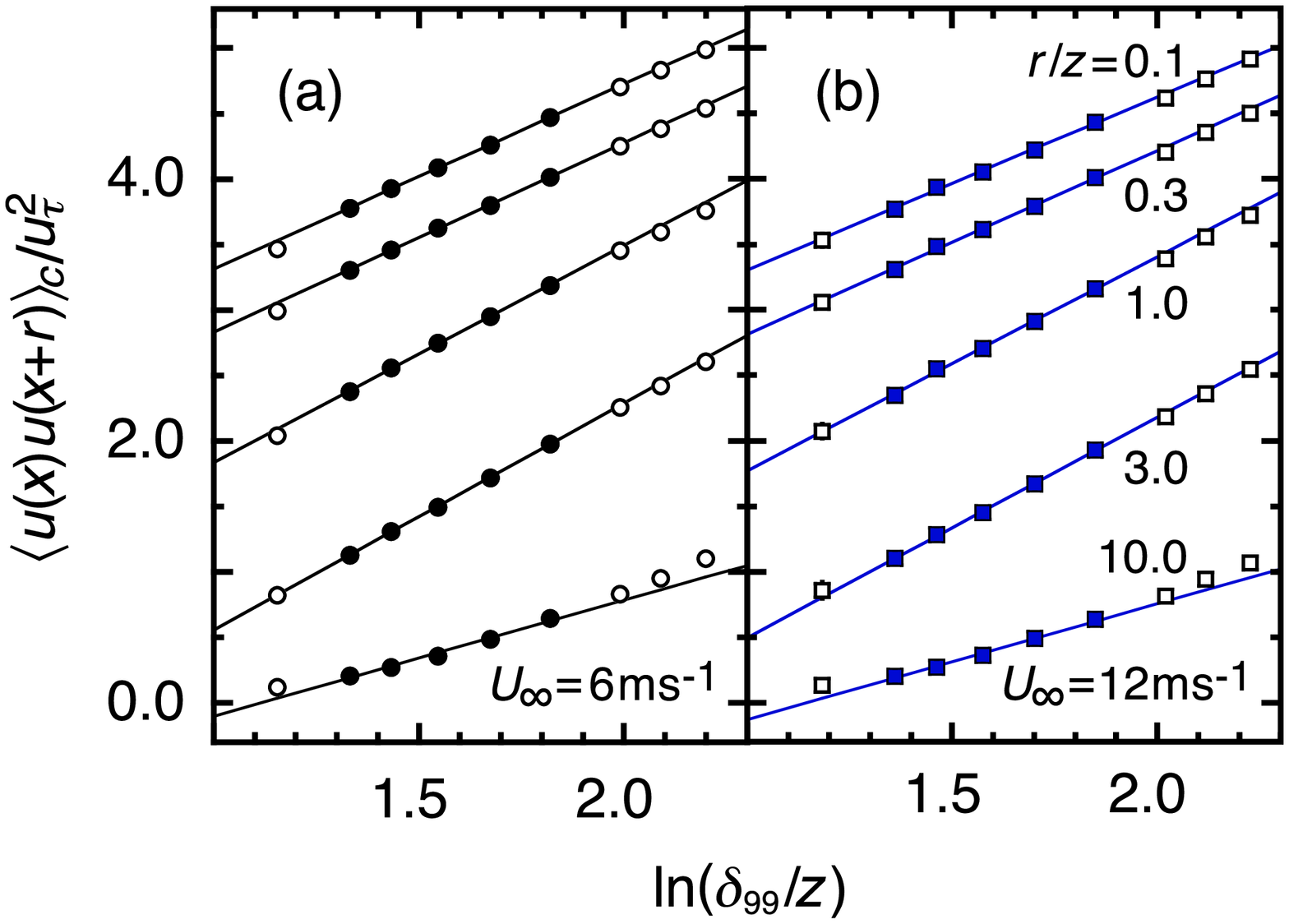}}
}
\caption{\label{f3} Two-point velocity cumulant $\langle u(x,z) u(x+r,z) \rangle_c/u_{\tau}^2$ $= \langle u(x,z) u(x+r,z) \rangle/u_{\tau}^2$ against $\ln (\delta_{99}/z )$ at $r/z = 0.1$, $0.3$, $1.0$, $3.0$, and $10.0$ for (a) $U_{\infty} = 6$\,m\,s$^{-1}$ and (b) $12$\,m\,s$^{-1}$. The filled symbols lie in the constant-flux sublayer. To these, solid lines are regression fits of Eq.~(\ref{eq3a}). Although we provide $\pm 2 \sigma$ errors on all the data in the same manner as in Figs.~\ref{f2}(a) and \ref{f2}(b), none of them are discernible. }
\end{figure} 
\begin{figure}[htbp]
\resizebox{7.84cm}{!}{\includegraphics*[3.0cm,1.7cm][17.5cm,28.4cm]{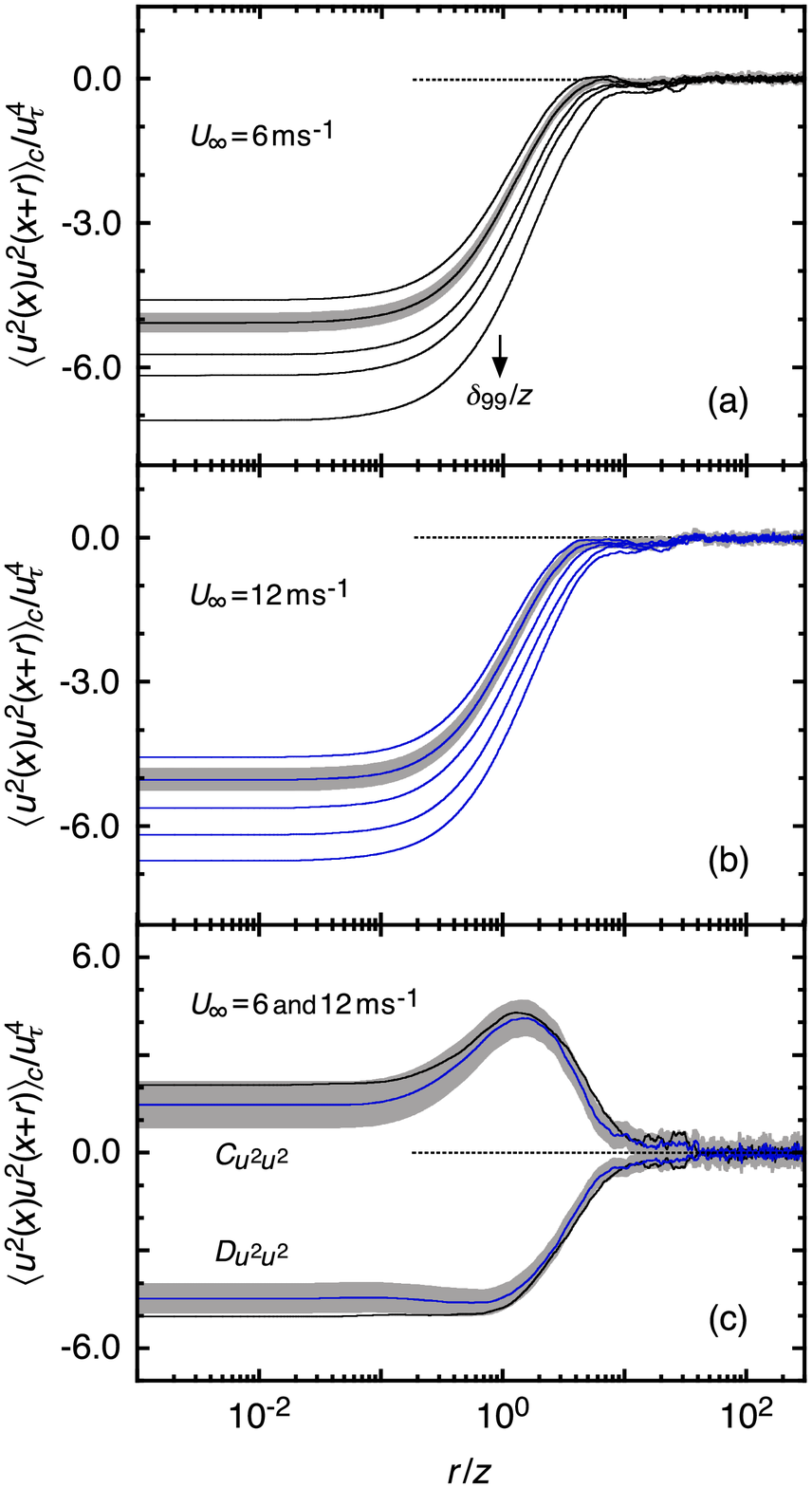}}
\caption{\label{f4} Same as in Fig.~\ref{f2} but for $\langle u^2(x,z) u^2(x+r,z) \rangle_c/u_{\tau}^4 = [\langle u^2(x,z) u^2(x+r,z) \rangle - \langle u(z)^2 \rangle^2 - 2 \langle u(x,z) u(x+r,z) \rangle^2]/u_{\tau}^4$. }
\end{figure} 
\begin{figure}[tbp]
\rotatebox{90}{
\resizebox{5.50cm}{!}{\includegraphics*[3.0cm,7.2cm][16.7cm,26.5cm]{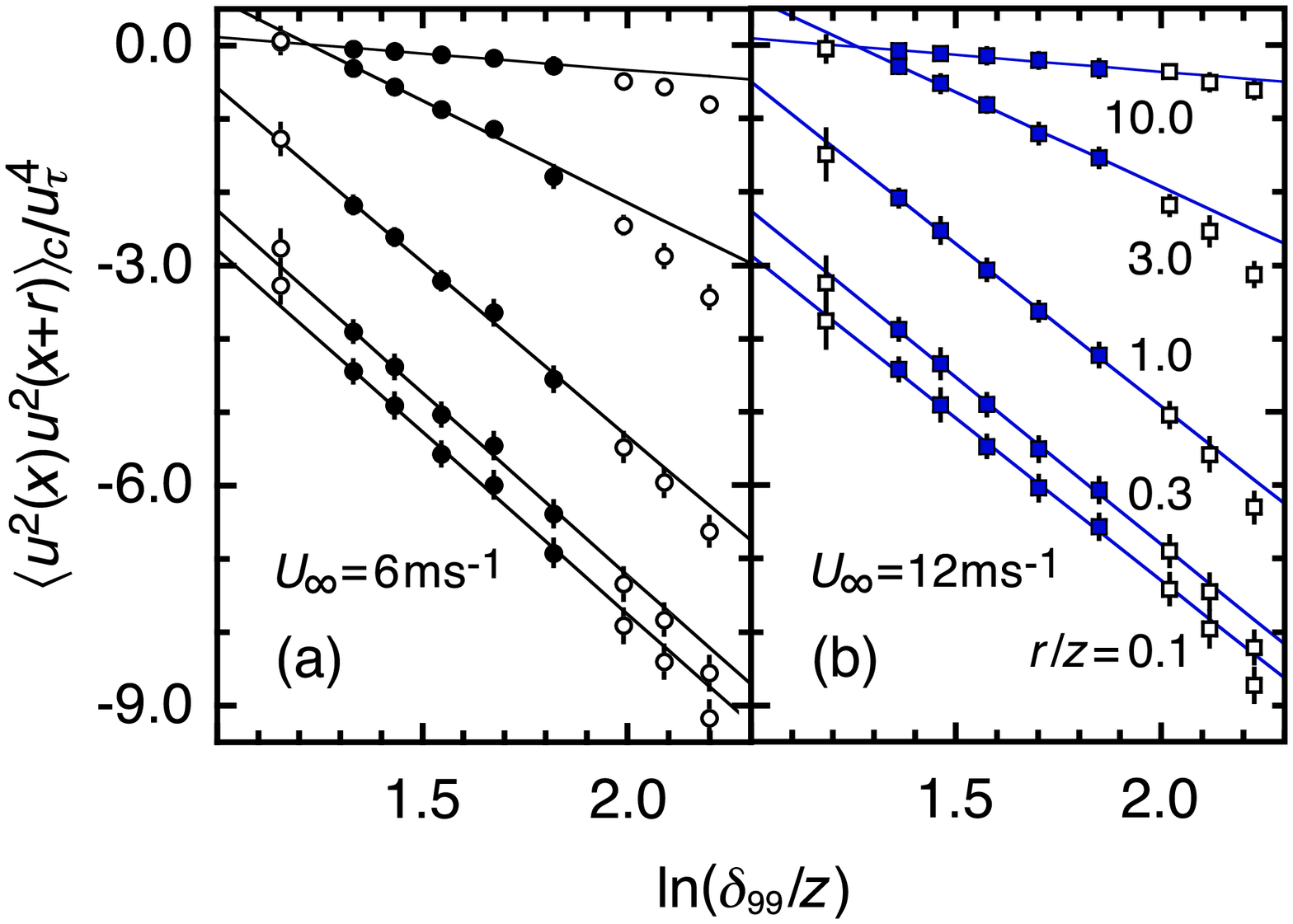}}
}
\caption{\label{f5} Same as in Fig.~\ref{f3} but for $\langle u^2(x,z) u^2(x+r,z) \rangle_c/u_{\tau}^4 = [\langle u^2(x,z) u^2(x+r,z) \rangle - \langle u(z)^2 \rangle^2 - 2 \langle u(x,z) u(x+r,z) \rangle^2]/u_{\tau}^4$. }
\end{figure} 
\begin{figure}[tbp]
\resizebox{7.84cm}{!}{\includegraphics*[3.0cm,1.7cm][17.5cm,28.4cm]{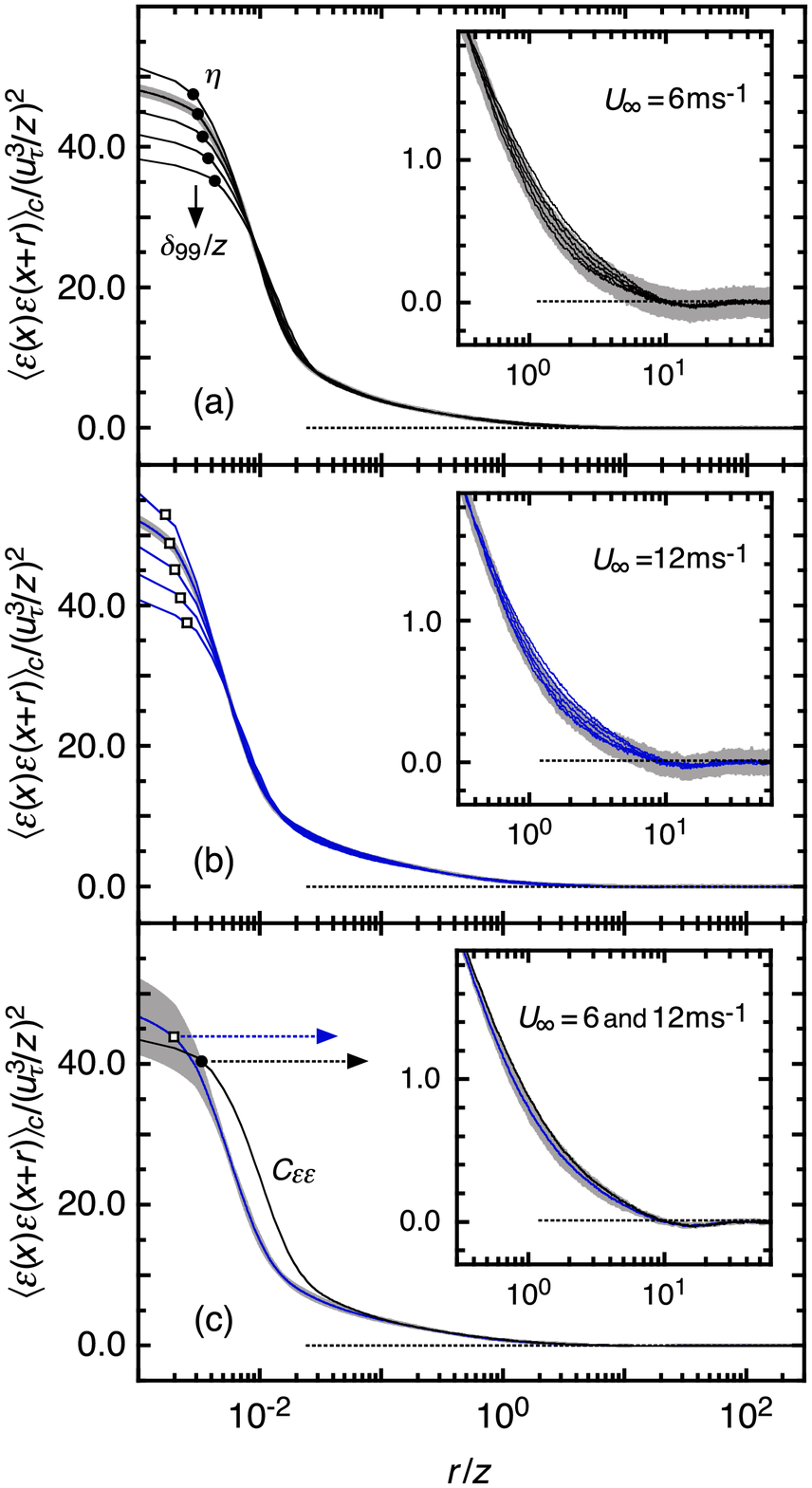}}
\caption{\label{f6} Two-point cumulant of the rate of the energy dissipation in the constant-flux sublayer $\langle \varepsilon (x,z) \varepsilon (x+r,z) \rangle_c/(u_{\tau}^3/z)^2$ $= [\langle  \varepsilon (x,z) \varepsilon (x+r,z) \rangle - \langle \varepsilon (z) \rangle^2 ]/(u_{\tau}^3/z)^2$ against $r/z$ for (a) $U_{\infty} = 6$\,m\,s$^{-1}$ and (b) $12$\,m\,s$^{-1}$. The solid arrow indicates an increase in $\delta_{99}/z$. The panel (c) shows $C_{\varepsilon\varepsilon}(r/z)$ of Eq.~(\ref{eq3b}). We provide examples of $\pm 2 \sigma$ errors as in Figs.~\ref{f2} and \ref{f4}. The circles and squares indicate the Kolmogorov length $\eta$. The insets are for large values of $r/z$.}
\end{figure} 

The inner bound of the sublayer depends on roughness of the wall \cite{t76, hvbs13}. Specifically for our roughness, we have observed its direct effect at $z \lesssim 40$\,mm, i.e., $\lesssim 0.10 \delta_{99}$, by shifting measurement positions slightly in the $x$ direction. The outer bound depends on the turbulence itself. Our estimate of $z/\delta_{99} \simeq 0.28$ is larger than those of $0.15$ in most studies \cite{mmhs13, hvbs13, vhs15, mm19}. Since the sublayer at any finite Reynolds number is an approximation (Sec.~\ref{S1}), no unique definition exists about its bound. An estimate similar to ours is actually found in the literature \cite{ofsbta17}.

We have used $\langle -uw \rangle^{1/2}$ as the friction velocity $u_{\tau}$. The results for the von K\'arm\'an constant $\kappa$ in Table~\ref{t1} are small with respect to the standard value of $\kappa = 0.39 \pm 0.02$ \cite{mmhs13}. Since it yet lies within $\pm 2\sigma$ errors of our results, we have adopted $\pm 2\sigma$ as a typical level of the uncertainties. Then, $c_{u^2} = C_{u^2}(0)$ and $d_{u^2} = D_{u^2}(0)$ in Table~\ref{t1} are consistent with those in Sec.~\ref{S1} summarized from the literature \cite{mmhs13, hvbs13, vhs15, ofsbta17, smhffhs18, hvbs12}.

The logarithmic scaling is also observed in Fig.~\ref{f1}(d) for $\langle u^4 \rangle_c = \langle u^4 \rangle - 3 \langle u^2 \rangle^2$. We expect this from Eq.~(\ref{eq3a}) at $r = 0$ and $n + m = 4$. The results for $C_{u^4}(0)$ and $D_{u^4}(0) $ in Table \ref{t1} are consistent within $\pm 2\sigma$ errors between the cases of $U_{\infty} = 6$ and $12$\,m\,s$^{-1}$ (see also Appendix \ref{apA}).

We have $\langle u^4 \rangle_c = \langle u^4 \rangle - 3 \langle u^2 \rangle^2 < 0$ in the constant-flux sublayer. Its streamwise fluctuations $u$ are known to be sub-Gaussian, i.e., $\langle u^4 \rangle < 3 \langle u^2 \rangle^2$. The spanwise and the wall-normal fluctuations are super-Gaussian \cite{ff96}.

The present and some other data exhibit $\langle u^4 \rangle_c / \langle u^2 \rangle_c^2 \simeq -0.3$ \cite{hvbs13, vhs15, smhffhs18, ff96, mm13}. Since $\langle u^4 \rangle_c$ is not so significant, $\langle u^4 \rangle = \langle u^4 \rangle_c + 3 \langle u^2 \rangle_c^2$ could be approximated by $3\langle u^2 \rangle_c^2 = 3\langle u^2 \rangle^2$, to which Eq.~(\ref{eq1b}) for $\langle u^2 \rangle$ is applicable \cite{mm13}. Nevertheless, the exact law is Eq.~(\ref{eq3a}) for the cumulant $\langle u^4 \rangle_c$.

Finally in Fig.~\ref{f1}(e), the ratio of $\langle \varepsilon \rangle_c = \langle \varepsilon \rangle$ to $u_{\tau}^3/z$ lies at around $1/\kappa = 2.6$ (dotted lines), which corresponds to the standard value of $\kappa = 0.39$ \cite{mmhs13}. Although that ratio appears to vary in contrast to the law of Eq.~(\ref{eq1c}) or (\ref{eq3b}) \cite{lm15}, a further study is desired because $\langle \varepsilon (x, z) \varepsilon (x+r, z) \rangle_c$ satisfies Eq.~(\ref{eq3b}) in a range of separations $r$ (see below and also Appendix \ref{apB}).

To confirm that the spatial and temporal resolutions of our experiments were high enough for an estimation of $\varepsilon$, we consider $\langle (\partial_x u)^4 \rangle / \langle (\partial_x u)^2 \rangle^2 \simeq \langle \varepsilon^2 \rangle / \langle \varepsilon \rangle^2$. As for $U_{\infty} = 6$ and $12$\,m\,s$^{-1}$, its values in the constant-flux sublayer are $7.7$--$7.8$ and $8.5$--$8.6$. Given the microscale Reynolds numbers $\langle u^2 \rangle / \nu  \langle (\partial_x u)^2 \rangle^{1/2}$ of $400$--$410$ and $570$--$580$, they are consistent with results of the previous studies \cite{sa97}.

We are now to study scaling laws of two-point statistics for Eq.~(\ref{eq3a}) at $n = m = 1$ and $2$ as well as for Eq.~(\ref{eq3b}) at $n = m = 1$. These statistics are shown in Figs.~\ref{f2}--\ref{f6} as a function of $r/z$ or $\delta/z = \delta_{99}/z$.

Figure \ref{f2} shows $\langle u(x,z) u(x+r,z) \rangle_c/u_{\tau}^2$ against $r/z$ at several wall-normal distances $z$ lying in the constant-flux sublayer. Here $\langle u(x,z) u(x+r,z) \rangle_c = \langle u(x,z) u(x+r,z) \rangle$ is a velocity correlation that would offer basic information about the wall turbulence. While it is usual to study this correlation as a function of $r/(\nu/u_{\tau})$ \cite{ghhlm05} or $r/\delta$ \cite{mswc07, dn11a}, we have adopted $r/z$ from Eq.~(\ref{eq3a}).

If $r/z$ is less than a few times $10^{-2}$, $\langle u(x,z) u(x+r,z) \rangle_c$ at each of $\delta_{99}/z$ is almost constant at about the value of $\langle u^2(z) \rangle_c$. With an increase in $r/z$, it decays toward $0$. It is no longer persistent if $r/z$ exceeds a few times $10^1$.

If $r/z$ is fixed, $\langle u(x,z) u(x+r,z) \rangle_c$ becomes large with an increase in $\delta_{99}/z$ (an arrow). This is due to the logarithmic law of Eq.~(\ref{eq3a}). Actually in Fig.~\ref{f3}, data points of $\langle u(x,z) u(x+r,z) \rangle_c$ make up linear functions of $\ln (\delta_{99}/z)$. The same has been observed in a laser Doppler anemometry of a similar flow \cite{mizuno18}. By fitting Eq.~(\ref{eq3a}) to our data \cite{br03}, we calculate the functions $C_{uu}(r/z)$ and $D_{uu}(r/z)$. Those in Fig.~\ref{f2}(c) for $U_{\infty} = 6$ and $12$\,m\,s$^{-1}$ collapse individually to single curves,  the shapes of which are to be discussed in Sec.~\ref{S5}.

Figure \ref{f4} shows $\langle u^2(x,z) u^2(x+r,z) \rangle_c/u_{\tau}^4$ against $r/z$. Since $\langle u^2(x,z) u^2(x+r,z) \rangle_c = \langle u^2(x,z) u^2(x+r,z) \rangle - \langle u(z)^2 \rangle^2 - 2 \langle u(x,z) u(x+r,z) \rangle^2$ at $r=0$ is reduced to $\langle u^4(z) \rangle_c = \langle u^4(z) \rangle - 3 \langle u^2(z) \rangle^2$ in Fig.~\ref{f1}(d), this is a correlation of some non-Gaussian component of the velocity fluctuations $u$.

The two-point cumulant $\langle u^2(x,z) u^2(x+r,z) \rangle_c$ at each of $\delta_{99}/z$ is constant up to $r/z \simeq 10^{-1}$ and is persistent up to $r/z \simeq 10^1$. If $r/z$ is fixed, it is dependent on $\delta_{99}/z$. The reason is the logarithmic law of Eq.~(\ref{eq3a}) as confirmed in Fig.~\ref{f5}. Its functions $C_{u^2u^2}(r/z)$ and $D_{u^2u^2}(r/z)$ in Fig.~\ref{f4}(c) are consistent between the cases of $U_{\infty} = 6$ and $12$\,m\,s$^{-1}$. These are analogous to results for $\langle u(x,z) u(x+r,z) \rangle_c$ in Figs.~\ref{f2} and \ref{f3}, although the shapes of the curves are entirely different.

Figure \ref{f6} shows $\langle \varepsilon (x,z) \varepsilon (x+r,z) \rangle_c/(u_{\tau}^3/z)^2$ against $r/z$. Here $\langle \varepsilon (x,z) \varepsilon (x+r,z) \rangle_c = \langle  \varepsilon (x,z) \varepsilon (x+r,z) \rangle - \langle \varepsilon (z) \rangle^2$ is a correlation of the dissipation rate $\varepsilon$. While it is usual to study this correlation as a function of $r/\eta$ \cite{cgs03, po97, mthk06}, we have adopted $r/z$ from Eq.~(\ref{eq3b}).

There is an enhancement of $\langle \varepsilon (x,z) \varepsilon (x+r,z) \rangle_c$ at the smallest separations $r$ \cite{cgs03, po97, mthk06}. They lie in the dissipative range, which extends from the Kolmogorov length $\eta$ (circles or squares) by a factor of $20$--$30$ (dotted arrows). Since the fluid viscosity $\nu$ is not negligible, Eq.~(\ref{eq3b}) does not hold there. The enhancement of $\langle \varepsilon (x,z) \varepsilon (x+r,z) \rangle_c$ is to be described rather by the small-scale intermittency \cite{igk09, sa97}.

Above that dissipative range, there exist the inertial and the energy-containing ranges. Throughout these two, $\langle \varepsilon (x,z) \varepsilon (x+r,z) \rangle_c/(u_{\tau}^3/z)^2$ is independent of $\delta_{99}/z$ in accordance with Eq.~(\ref{eq3b}). It persists up to a large separation \cite{cgs03, po97, mthk06}, i.e., $r/z \simeq 10^1$, as predicted originally by Landau \cite{ll59, mthk06}. We calculate the average at each of $r/z$ to obtain the function $C_{\varepsilon \varepsilon}(r/z)$. Those in Fig.~\ref{f6}(c) for $U_{\infty} = 6$ and $12$ m\,s$^{-1}$ collapse to a single curve at least at separations $r$ in the inertial and the energy-containing ranges.

\section{Discussion} \label{S5}

Having confirmed the scaling laws of Eq.~(\ref{eq3}), we discuss their functions $C_{uu}$, $D_{uu}$, $C_{u^2u^2}$, $D_{u^2u^2}$, and $C_{\varepsilon \varepsilon}$ in terms of the attached-eddy hypothesis \cite{t76,m17}.

\subsection{Scaling laws from attached eddies} \label{S5a}

\begin{figure}[tbp]
\rotatebox{90}{
\resizebox{5cm}{!}{\includegraphics*[5.4cm,3.8cm][18.6cm,26.7cm]{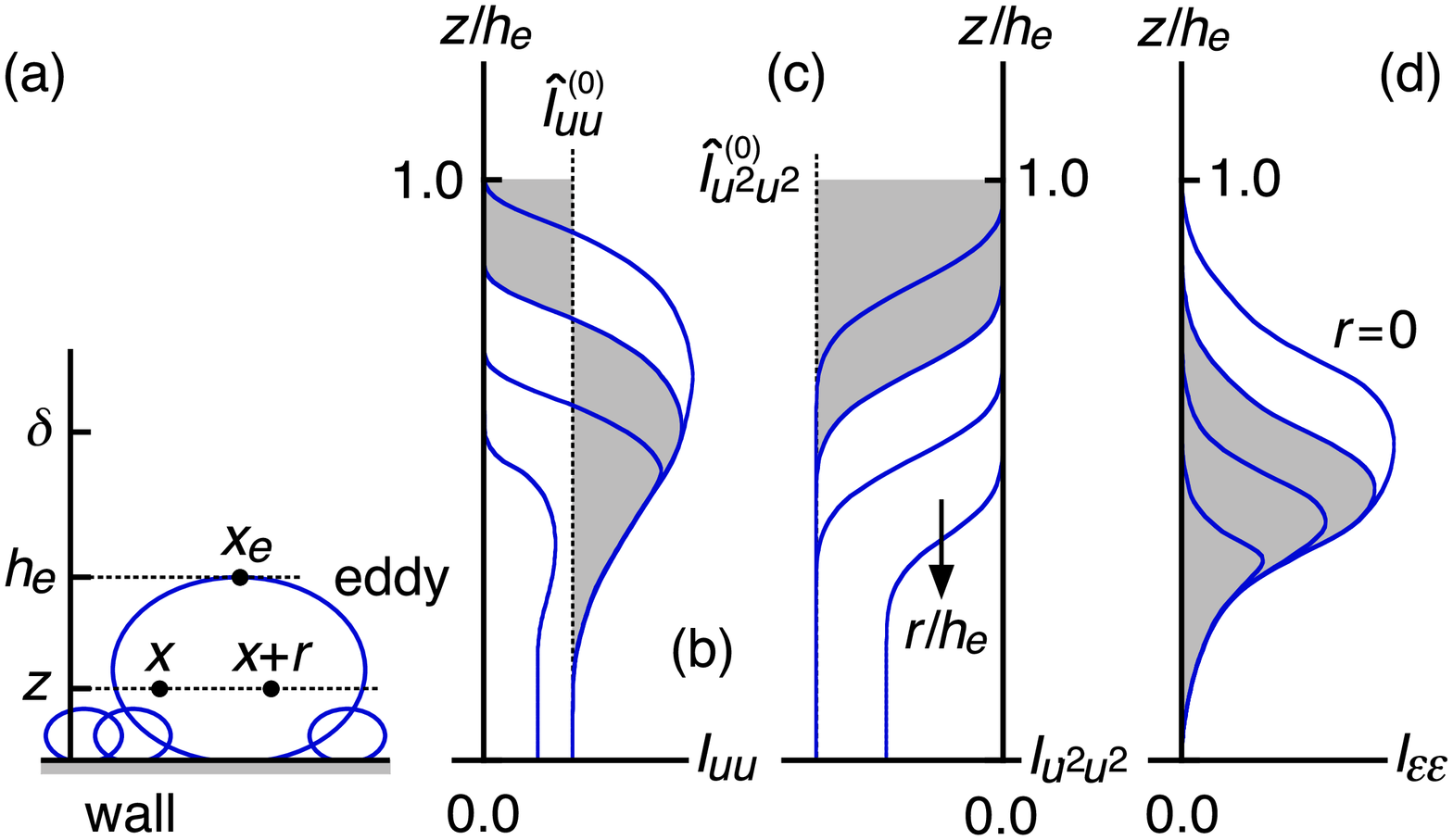}}
}
\caption{\label{f7} Schematic of (a) attached eddies, (b) $I_{uu}(r/h_e, z/h_e)$,  (c) $I_{u^2u^2}(r/h_e, z/h_e)$, and  (d) $I_{\varepsilon\varepsilon}(r/h_e, z/h_e)$. The solid arrow indicates an increase in $r/h_e$. The dotted lines are examples of $\hat{I}^{(0)}_{u^nu^n} (r/z)\varpropto D_{u^nu^n} (r/z)$ in Eq.~(\ref{eq12c}). They are to reproduce $D_{uu} \ge 0$ and $D_{u^2u^2} \le 0$ in Figs.~\ref{f2}(c) and \ref{f4}(c). As for integrations of $I_{u^nu^n}-\hat{I}^{(0)}_{u^nu^n}$ in Eq.~(\ref{eq12b}) and of $I_{\varepsilon\varepsilon}$ in Eq.~(\ref{eq14b}), the gray areas are examples to reproduce $C_{uu} (r/z)$, $C_{u^2u^2} (r/z)$, and $C_{\varepsilon \varepsilon} (r/z)$ in Figs. \ref{f2}(c), \ref{f4}(c), and \ref{f6}(c).}
\end{figure}

Figure \ref{f7}(a) is a schematic of the attached eddies. They have various finite sizes, a common shape that is extending from the wall, and a common characteristic velocity $u_{\tau}$. If $\mbox{\boldmath{$x$}}_{\rm e} = (x_{\rm e}, y_{\rm e}, h_{\rm e})$ is the most extended position of an eddy, $h_e$ is used as its size. It induces the streamwise velocity $u_e$ and the energy dissipation $\varepsilon_e$ at any position $\mbox{\boldmath{$x$}} = (x, y, z)$ as
\begin{equation}
\label{eq10}
\frac{u_{\rm e}(\mbox{\boldmath{$x$}})}{u_{\tau}}                 = f_u             \! \left( \!  \frac{\mbox{\boldmath{$x$}} - \mbox{\boldmath{$x$}}_{\rm e}}{h_{\rm e}} \! \right) \!
\ \ \mbox{and} \ \
\frac{\varepsilon_{\rm e}(\mbox{\boldmath{$x$}})}{u_{\tau}^3/h_e} = f_{\varepsilon} \! \left(  \! \frac{\mbox{\boldmath{$x$}} - \mbox{\boldmath{$x$}}_{\rm e}}{h_{\rm e}} \! \right) \! .
\end{equation}
Here $f_u$ and $f_{\varepsilon}$ are nondimensional functions that take finite nonzero values in a finite volume. As for the velocity function $f_u$, we impose a free-slip wall condition, i.e., $f_u \ne 0$ at $z = 0$.

The number of eddies of size $h_e$ per unit area of the wall is $N_e h_e^{-3} dh_e$  \cite{t76}. Here $N_e$ is a constant. Apart from $N_e$ and $h_e$, no quantity affects such a number density. Upon the wall, the distribution of the eddies is random and independent. They could overlap one another because we use them not as realistic organized structures but only as bases to describe the flow \cite{t76, mm19}.

The entire flow is a superposition of the eddies. Since they are random and independent, a sum of their cumulants is identical to the cumulant of this flow. The law for $z/\delta \rightarrow 0$ is used as a law for the constant-flux sublayer, in accordance with its asymptotic character (Sec.~\ref{S1}, see also Appendix \ref{apB}).

From this attached-eddy hypothesis, Eq.~(\ref{eq3}) is derived by simplifying the original derivation \cite{m17}. We again use nondimensional functions of nondimensional parameters such as $r/z$.

The velocity cumulant $\langle u^n(x,z) u^m(x+r,z) \rangle_c$ at any distance $z \le \delta$ is given by an integration from $h_e = z$ to $\delta$, 
\begin{subequations} \label{eq11}
\begin{equation} \label{eq11a}
\frac{\langle u^n(x,z)u^m(x+r,z) \rangle_c}{u_{\tau}^{n+m}} = \! N_e \! \! \int^{\delta}_z \! \! \frac{dh_e}{h_e} I_{u^nu^m} \! \! \left( \! \frac{r}{h_e}, \frac{z}{h_e} \! \right) \! .
\end{equation}
Here $I_{u^n u^m}$ is a cumulant of eddies of particular size $h_e$. As in the case of Eq.~(\ref{eq2}), it is related to moments $J_{u^l u^k}$ and $J_{u^l} = J_{u^l u^0}$. For example,
\begin{equation} \label{eq11b} 
I_{uu} \! \! \left( \! \frac{r}{h_e}, \frac{z}{h_e} \! \right) \! = \! J_{uu} \! \left( \! \frac{r}{h_e}, \frac{z}{h_e} \! \right) \! - J_{u}^2 \! \left( \! 0, \frac{z}{h_e} \! \right) \!.
\end{equation}
We use $\mbox{\boldmath{$r$}} = (r,0,0)$ to define $J_{u^lu^k}$ via an integration at that distance $z$,
\begin{equation} \label{eq11c}
J_{u^lu^k} \! \! \left( \! \frac{r}{h_e}, \frac{z}{h_e} \! \right) 
\! = \! \!
\iint \! \frac{dx_e}{h_e} \frac{dy_e}{h_e} f_u^l \! \! \left( \! \frac{\mbox{\boldmath{$x$}}                         - \mbox{\boldmath{$x$}}_e}{h_e} \! \right) \!
                                           f_u^k \! \! \left( \! \frac{\mbox{\boldmath{$x$}} + \mbox{\boldmath{$r$}} - \mbox{\boldmath{$x$}}_e}{h_e} \! \right) \! \! . 
\end{equation}
\end{subequations}
The first-order moment $J_u$ is not necessarily equal to $0$ \cite{pc82}. Although this was neglected in the original studies \cite{t76,m17}, their final results hold without any correction.

By using $\zeta = z/h_e$ and hence $d \zeta /\zeta = -dh_e /h_e$, we rewrite Eq.~(\ref{eq11a}) as a function of $z/\delta$ and $r/z$,
\begin{subequations} \label{eq12}
\begin{equation} \label{eq12a}
\! \frac{\langle u^n(x,z)u^m(x+r,z) \rangle_c}{u_{\tau}^{n+m}} = N_e \! \! \int^{1}_{z/\delta} \! \frac{d \zeta}{\zeta} I_{u^nu^m} \! \! \left( \frac{r}{z} \zeta, \zeta \right) \! . \! \!
\end{equation}
The free-slip wall condition implies that $I_{u^nu^m}(r \zeta /z, \zeta )$ at $ \zeta \rightarrow 0$ is not necessarily equal to $0$.  There is a function $\hat{I}^{(0)}_{u^nu^m}(r/z)$ such that $I_{u^nu^m} \rightarrow \hat{I}^{(0)}_{u^nu^m}$. By rewriting $I_{u^nu^m}$ as $I_{u^nu^m} - \hat{I}^{(0)}_{u^nu^m} + \hat{I}^{(0)}_{u^nu^m}$ in Eq.~(\ref{eq12a}) and by taking the limit $z/ \delta \rightarrow 0$, we derive Eq.~(\ref{eq3a}). Its functions $C_{u^nu^m}$ and $D_{u^nu^m}$ are
\begin{align}
\label{eq12b} &C_{u^nu^m} \! \! \left( \frac{r}{z} \right) \! 
 = N_e \! \! \int^1_0 \! \frac{d \zeta}{\zeta} \! \left[ I_{u^n u^m} \! \! \left( \frac{r}{z} \zeta, \zeta \right) \! - \hat{I}^{(0)}_{u^nu^m} \! \! \left( \frac{r}{z} \right) \right] \! , \\
\label{eq12c} &D_{u^nu^m} \! \! \left( \frac{r}{z} \right) \!
 = N_e                                                                                                                  \hat{I}^{(0)}_{u^nu^m} \! \! \left( \frac{r}{z} \right)         \! .
\end{align}
\end{subequations}
Because of $l \ge 1$ for $I_{u^nu^m} (r \zeta /z, \zeta ) - \hat{I}^{(0)}_{u^nu^m}(r/z) \varpropto \zeta^l$ at each of $r/z$ in the limit $\zeta \rightarrow 0$, the integration for $C_{u^nu^m}$ in Eq.~(\ref{eq12b}) is convergent. The divergent component has been removed as $D_{u^nu^m}$ in Eq.~(\ref{eq12c}). Likewise, at $n=1$ and $m=0$, Eq.~(\ref{eq12}) could yield the law of Eq.~(\ref{eq1a}) for the mean velocity $U(z)$ \cite{pc82},

The functional form of Eq.~(\ref{eq3a}) implies that it is always affected by the turbulence thickness $\delta$. Such an effect is through largest eddies of $h_e = \delta$, which contribute to all of nonzero cumulants \cite{m19}.

For the local rate of the energy dissipation $\varepsilon$, its cumulant is given by
\begin{subequations} \label{eq13}
\begin{equation} \label{eq13a}
\! \frac{\langle \varepsilon^n(x,z) \varepsilon^m(x+r,z) \rangle_c}{(u_{\tau}^3/z)^{n+m}} \! 
= 
N_e \! \! \int^{\delta}_z \! \! \frac{dh_e}{h_e} I_{\varepsilon^n \varepsilon^m} \! \! \left( \! \frac{r}{h_e}, \frac{z}{h_e} \! \right) \! . \! \!
\end{equation}
Here $I_{\varepsilon^n \varepsilon^m}$ is a cumulant of eddies of size $h_e$ that has been multiplied by a factor of $(z/h_e)^{n+m}$. The cumulant itself is related to moments $J_{\varepsilon^l \varepsilon^k}$ and  $J_{\varepsilon^l} = J_{\varepsilon^l \varepsilon^0}$. For example,
\begin{equation} \label{eq13b}
I_{\varepsilon \varepsilon} \! \! \left( \! \frac{r}{h_e}, \frac{z}{h_e} \! \right) \! = \frac{z^{2}}{h_e^{2}} \left[ J_{\varepsilon \varepsilon} \! \! \left( \! \frac{r}{h_e}, \frac{z}{h_e} \! \right) \! 
                                                                                                                    - J_{\varepsilon}^2           \! \! \left( \!             0, \frac{z}{h_e} \! \right) \! \right] .
\end{equation}
The moment $J_{\varepsilon^l \varepsilon^k}$ is defined as
\begin{equation} \label{eq13c}
J_{\varepsilon^l \varepsilon^k} \! \! \left( \! \frac{r}{h_e}, \frac{z}{h_e} \! \right) 
\! = \! \!
\iint \! \frac{dx_e}{h_e} \frac{dy_e}{h_e} f_{\varepsilon}^l \! \! \left( \! \frac{\mbox{\boldmath{$x$}}                         - \mbox{\boldmath{$x$}}_e}{h_e} \! \right) \!
                                           f_{\varepsilon}^k \! \! \left( \! \frac{\mbox{\boldmath{$x$}} + \mbox{\boldmath{$r$}} - \mbox{\boldmath{$x$}}_e}{h_e} \! \right) \! \! .     
\end{equation}
\end{subequations}
We use $\zeta = z/h_e$ to rewrite Eq.~(\ref{eq13a}),
\begin{subequations} \label{eq14}
\begin{equation} \label{eq14a}
\frac{\langle \varepsilon^n(x,z)\varepsilon^m(x+r,z) \rangle_c}{(u_{\tau}^3/z)^{n+m}} 
 = 
 N_e \! \! \int^{1}_{z/\delta} \! \frac{d \zeta}{\zeta} I_{\varepsilon^n \varepsilon^m} \! \! \left( \frac{r}{z} \zeta, \zeta \right) \! .
\end{equation}
Since $I_{\varepsilon^n \varepsilon^m}$ has a factor of $(z/h_e)^{n+m} = \zeta^{n+m}$ with $n+m \ge 1$ as in the case of Eq.~(\ref{eq13b}), the integration of Eq. (\ref{eq14a}) is convergent even in the limit $z/\delta \rightarrow 0$. Thus, Eq. (\ref{eq14a}) yields Eq.~(\ref{eq3b}). Its function $C_{\varepsilon^n \varepsilon^m}$ is
\begin{equation} \label{eq14b}
C_{\varepsilon^n \varepsilon^m} \! \! \left( \frac{r}{z} \right) = N_e \! \! \int^1_0 \! \frac{d \zeta}{\zeta}I_{\varepsilon^n \varepsilon^m} \! \! \left( \frac{r}{z} \zeta, \zeta \right) \! .
\end{equation}
\end{subequations}
That at $n=1$ and $m=0$ corresponds to the law of Eq. (\ref{eq1c}) for the average $\langle \varepsilon (z) \rangle$.

\subsection{Implication for attached eddies} \label{S5b}

By utilizing $D_{uu}(0) = D_{u^2}(0)$ and $D_{u^2u^2}(0) = D_{u^4}(0)$, we constrain the value of $N_e$. From Eq.~(\ref{eq12c}),
\begin{subequations} \label{eq15}
\begin{equation} \label{eq15a}
\frac{D_{u^4}(0)}{N_e} + \frac{2D_{u^2}^2(0)}{N_e^2} = \hat{I}^{(0)}_{u^4}(0) + 2 \! \left[ \hat{I}^{(0)}_{u^2}(0) \right]^2 \! \ge 0.
\end{equation}
This is because $\hat{I}^{(0)}_{u^4}(0)$ and $\hat{I}^{(0)}_{u^2}(0)$ are cumulants of orders $4$ and $2$, i.e.,  $\langle \alpha^4 \rangle_c + 2 \langle \alpha^2 \rangle_c^2 = \langle (\alpha - \langle \alpha \rangle)^4 \rangle - \langle (\alpha - \langle \alpha \rangle)^2 \rangle^2 = \langle [ (\alpha - \langle \alpha \rangle)^2 - \langle (\alpha - \langle \alpha \rangle)^2 \rangle ]^2 \rangle \ge 0$. Then,
\begin{equation} \label{eq15b}
N_e \le - \frac{2 D_{u^2}^2(0)}{D_{u^4}(0)}
\ \ \mbox{if} \ \
D_{u^4}(0) < 0 .
\end{equation}
\end{subequations}
The results for $D_{u^2}(0)$ and $D_{u^4}(0)$ in Table \ref{t1} yield $N_e \lesssim 0.6$. Here $N_e$ serves as a number of attached eddies of size $h_e$ per each volume of $h_e^3$. It is thereby inferred that the eddies are rather sparse so that those of similar sizes do not  overlap one another, although each of them contains many smaller eddies.

To explain $C_{uu}$, $D_{uu}$, $C_{u^2u^2}$, $D_{u^2u^2}$ and $C_{\varepsilon\varepsilon}$ observed in Figs. \ref{f2}(c), \ref{f4}(c), and \ref{f6}(c), we discuss the likely shapes of $I_{uu}$, $I_{u^2u^2}$, and $I_{\varepsilon\varepsilon}$. They are illustrated schematically in Figs. \ref{f7}(b)--\ref{f7}(d).

The functions $C_{uu}$, $C_{u^2u^2}$ and $C_{\varepsilon\varepsilon}$ in Eqs.~(\ref{eq12b}) and (\ref{eq14b}) are due to eddies of wall-normal sizes $h_e$ that are comparable to the observing distance $ z$ \cite{t76}. Since those in Figs.~\ref{f2}(c), \ref{f4}(c), and \ref{f6}(c) are persistent up to almost the same separation of $r/z \simeq 10^1$, the streamwise to wall-normal size ratio of an eddy is about $10^1$.

The functions  $D_{uu}$ and $D_{u^2u^2}$ in Eq.~(\ref{eq12c}) are due to wall-adjacent portions of the eddies \cite{t76}. Since those in Figs.~\ref{f2}(c) and \ref{f4}(c) are again persistent up to $r/z \simeq 10^1$, the streamwise size of such a portion is proportional to the distance $z$.

From the shapes of $C_{u^2u^2}$ and  $D_{u^2u^2}$ observed in Fig. \ref{f4}(c), we expect $I_{u^2u^2} < 0$ for all pairs of $r/h_e$ and $z/h_e$ (see Fig.~\ref{f7}). The sub-Gaussianity of the fluctuations $u$, i.e., $\langle u^4 \rangle_c < 0$ in Fig.~\ref{f1}(d), extends over the entire ranges of streamwise separations $r$ and of wall-normal distances $z$ in each of the eddies.

The function $C_{\varepsilon\varepsilon}$ in Fig.~\ref{f6}(c) differs between the cases of $U_{\infty} = 6$ and $12$\,m\,s$^{-1}$ at $r \lesssim 30 \eta$ (dotted arrows). This dissipative range is where the attached-eddy hypothesis holds no longer \cite{m17, mm19}. From Eq.~(\ref{eq10}), the Kolmogorov length $\eta$ is obtained as $(\nu^3/\varepsilon_e)^{1/4} \varpropto (\nu /u_{\tau})^{3/4} h_e^{1/4}$, which is not proportional to the eddy size $h_e$. To describe motions in the dissipative range, we require some other class of small eddies, e.g., vortex tubes \cite{igk09, sa97, mhk07}, albeit not essential to statistics like $\langle u^n(x,z) u^m(x+r,z) \rangle_c$ that are almost constant there (Figs.~\ref{f2} and \ref{f4}).

\section{Concluding Remarks} \label{S6}

For the constant-flux sublayer of wall turbulence, the logarithmic scaling of $\langle u^n(x,z) u^m(x+r,z) \rangle_c$ in Eq.~(\ref{eq3a}) and the nonlogarithmic scaling of $\langle \varepsilon^n(x,z) \varepsilon^m(x+r,z) \rangle_c$ in Eq.~(\ref{eq3b}) have been studied experimentally. We have done experiments of boundary layers and have obtained those two-point cumulants at several distances $z$ from the wall (Sec.~\ref{S3}). The results in Figs.~\ref{f2}--\ref{f6} are consistent with Eq.~(\ref{eq3a}) at $n=m=1$ and $2$ as well as with Eq.~(\ref{eq3b}) at $n=m=1$ in the inertial and the energy-containing ranges of the separations $r$ (Sec.~\ref{S4}).

The mathematical reason for such a scaling law is that the constant-flux sublayer has only two local parameters, i.e., the distance $z$ and the friction velocity $u_{\tau}$. Under Galilean invariance, only laws for cumulants are extended systematically from the law of Eq.~(\ref{eq1a}) for the mean velocity $U(z)$. The logarithmic or nonlogarithmic character of the scaling is determined by a condition at the wall surface (Sec.~\ref{S2}). Even if those two points lie in the $y$ or $z$ direction \cite{m17}, we could rely on the same reasoning.

Having confirmed the scaling laws of Eq.~(\ref{eq3}), we have related them to the attached eddies of Townsend \cite{t76,m17}. Since the eddy number $N_e$ is rather small, eddies of each size $h_e$ do not overlap one another. The streamwise size of an eddy is about $10$ times its wall-normal size $h_e$. Within the individual eddies, the streamwise fluctuations $u$ are sub-Gaussian (Sec.~\ref{S5}).

These characteristics of the attached eddies might have been optimized to maximize the momentum flux $\rho \langle -uw \rangle$. Actually from recent applications of variational calculus to convective systems \cite{hcd14, mks18b}, it has been inferred that their scaling laws and eddy structures are optimal for their heat transfer. Such a study is equally desired for momentum transfer in wall turbulence.

We also desire to study cases other than the boundary layers. As noted in Sec.~\ref{S1}, among various configurations of the wall turbulence, $c_{u^2} = C_{u^2}(0)$ of Eq.~(\ref{eq1b}) is not common \cite{mmhs13, hvbs13, vhs15, ofsbta17, smhffhs18, hvbs12}. Then, $C_{u^nu^m}(r/z)$ of Eq.~(\ref{eq3a}) and possibly $C_{\varepsilon^n \varepsilon^m}(r/z)$ of Eq.~(\ref{eq3b}) are not universal. Nevertheless, since $d_{u^2} = D_{u^2}(0)$ is common, $D_{u^nu^m}(r/z)$ is expected to be universal. The reason would be that only $D_{u^nu^m}$ is determined locally in the constant-flux sublayer (Sec.~\ref{S2}).

Thus far, we have studied two-point cumulants, which are useful if wall turbulence is modeled as a superposition of finite-size motions like those of the attached eddies. Although it is usual to study energy spectra \cite{mm19}, their wavelengths do not correspond to any particular size of the individual motions \cite{m19}.

Finally, we remark on the turbulence thickness $\delta$. Wall turbulence is known to include long organized structures with streamwise lengths that exceed more than $10$ times the thickness $\delta$ \cite{mswc07, dn11b}. The ratio $r/\delta$ has accordingly been used as a parameter of two-point statistics \cite{mswc07, dn11a}. However, in our results and in some others \cite{m19, mizuno18}, those structures are not discernible. They are meandering so that their total lengths do not appear in one-dimensional correlations and spectra. Furthermore, the structures lie essentially away from the wall and do not affect any law of the constant-flux sublayer at least at $z/\delta \rightarrow 0$ \cite{mm19}. We have considered such asymptotic laws alone (see also Appendix \ref{apB}). Given the limit $z/\delta \rightarrow 0$, since the distance $z$ is always not equal to $0$, $r/\delta$ does not deserve to be a parameter. The true nondimensional parameters are $r/z$ and $\delta/z$ as demonstrated here for cases of the scaling laws of Eq.~(\ref{eq3}).

 \vspace{-3mm}

\begin{acknowledgments}
This work was supported in part by KAKENHI Grants No. 17K00526 and No. 19K03967.
\end{acknowledgments}

\begin{figure*}
\begin{center}
\rotatebox{90}{
\resizebox{5.50cm}{!}{\includegraphics*[9.9cm,1.9cm][18.5cm,29.2cm]{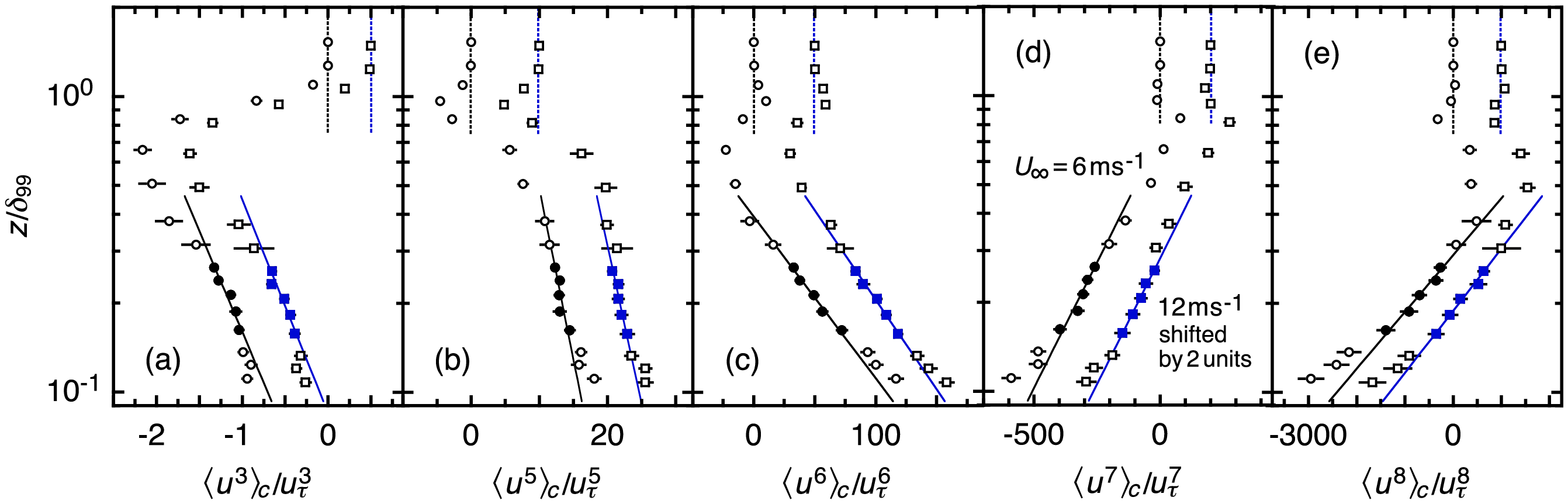}}
}
\caption{\label{f8} Same as in Fig.~\ref{f1} but for (a) $\langle u^3(z) \rangle_c = \langle u^3(z) \rangle$, (b) $\langle u^5(z) \rangle_c = \langle u^5(z) \rangle -10 \langle u^3(z) \rangle \langle u^2(z) \rangle$, (c) $\langle u^6(z) \rangle_c = \langle u^6(z) \rangle -15 \langle u^4(z) \rangle \langle u^2(z) \rangle -10 \langle u^3(z) \rangle^2 +30 \langle u^2(z) \rangle^3$, (d) $\langle u^7(z) \rangle_c = \langle u^7(z) \rangle -21 \langle u^5(z) \rangle \langle u^2(z) \rangle -35 \langle u^4(z) \rangle \langle u^3(z) \rangle +210 \langle u^3(z) \rangle \langle u^2(z) \rangle^2$, and (e) $\langle u^8(z) \rangle_c = \langle u^8(z) \rangle -28 \langle u^6(z) \rangle \langle u^2(z) \rangle -56 \langle u^5(z) \rangle \langle u^3(z) \rangle - 35 \langle u^4(z) \rangle^2 +420 \langle u^4(z) \rangle \langle u^2(z) \rangle^2 +560 \langle u^3(z) \rangle^2 \langle u^2(z) \rangle -630 \langle u^2(z) \rangle^4$ nondimensionalized with use of $u_{\tau}$. The solid lines are regression fits of Eq.~(\ref{eq3a}).
}
\end{center}
\end{figure*} 
\begin{table}[tbp]
\begingroup
\squeezetable
\caption{\label{t2} Parameters $C_{u^n}(0)$ and $D_{u^n}(0)$ of Eq.~(\ref{eq3a}) for $U_{\infty} = 6$ and $12$\,m\,s$^{-1}$. The uncertainties are $\pm 2 \sigma$ errors.}
\begin{ruledtabular}
\begin{tabular}{lcc}
                             & $U_{\infty} = 6$\,m\,s$^{-1}$ & $U_{\infty} = 12$\,m\,s$^{-1}$ \\  \hline
$C_{u^3}(0)$                 & $-2.17 \pm 0.33$              & $-1.99 \pm 0.34$               \\
$D_{u^3}(0)$                 & $0.64  \pm 0.13$              & $0.60  \pm 0.13$               \\
$C_{u^5}(0)$                 & $7.35  \pm 4.13$              & $5.35  \pm 5.11$               \\
$D_{u^5}(0)$                 & $3.75  \pm 1.58$              & $4.04  \pm 1.97$               \\
$C_{u^6}(0) = C_{u^3u^3}(0)$ & $-74.4 \pm 24.6$              & $-63.4 \pm 23.5$               \\
$D_{u^6}(0) = D_{u^3u^3}(0)$ & $79.5  \pm 9.8$               & $71.7  \pm 9.2$                \\
$C_{u^7}(0)$                 & $82    \pm 117$               & $122   \pm 151$                \\
$D_{u^7}(0)$                 & $-255  \pm 46$                & $-255  \pm 59$                 \\
$C_{u^8}(0) = C_{u^4u^4}(0)$ & $2790  \pm 950$               & $2460  \pm 910$                \\
$D_{u^8}(0) = D_{u^4u^4}(0)$ & $-2260 \pm 380$               & $-2070 \pm 350$                \\
\end{tabular}
\end{ruledtabular}
\endgroup
\end{table}

\appendix
\section{OTHER LOGARITHMIC LAWS} \label{apA}

 Figure \ref{f8} shows one-point velocity cumulants $\langle u^n \rangle_c$ at $n=3$ and $5$ to $8$ \cite{my71, ks77}. Those for $z/\delta_{99} \lesssim 0.10$ are not included because statistical errors are too large owing to the roughness of the wall surface (see Sec.~\ref{S4}). Within the constant-flux sublayer from $z/\delta_{99} \simeq 0.14$ to $0.28$ (filled symbols), we find that the cumulants obey the logarithmic laws of $\langle u^n(z) \rangle_c / u_{\tau}^n = C_{u^n}(0) + D_{u^n}(0) \ln (\delta_{99}/z)$, i.e., Eq.~(\ref{eq3a}) at $m=0$. Their parameters are summarized in Table \ref{t2}. Between the cases of $U_{\infty} = 6$ and $12$\,m\,s$^{-1}$, the values are consistent within $\pm 2\sigma$ errors.

These logarithmic laws offer mutually independent information about non-Gaussianity of the turbulence. If it were exactly Gaussian, we would have $\langle u^n \rangle_c \equiv 0$ at $n \ge 3$ \cite{my71, ks77}. The same scaling is expected for the corresponding two-point cumulants, e.g., $\langle u^3(x,z) u^3(x+r,z) \rangle_c$ and  $\langle u^4(x,z) u^4(x+r,z) \rangle_c$.

\section{ASYMPTOTIC EXPANSION} \label{apB}

Throughout our study, we have taken the limit $z/\delta \rightarrow 0$ and thereby invoked a sublayer where the momentum flux $\rho \langle -uw \rangle$ is constant at a value of $\rho u_{\tau}^2$. This constant-flux sublayer is yet wide in units of $\nu/u_{\tau}$, i.e., $\nu /u_{\tau} \ll z \ll \delta$, at a high Reynolds number $\delta u_{\tau} / \nu \gg 1$.

If the limit is not taken, $\langle -uw \rangle$ is not a constant. For example, in pipes and channels, $\langle -uw \rangle$ varies linearly with $z/\delta$ \cite{my71}. A similar but nonlinear variation is expected for boundary layers. We are to show that the logarithmic law of Eq.~(\ref{eq3a}) and the nonlogarithmic law of Eq.~(\ref{eq3b}) are not exact there. That is, for a study of such a scaling law, the constant-flux sublayer in the limit $z/\delta \rightarrow 0$ is an essential notion.

First, we consider $\langle uw(z) \rangle_c /u_{\tau}^2 = \langle uw(z) \rangle /u_{\tau}^2$. The prediction of the attached-eddy hypothesis is like those in Sec.~\ref{S5},
\begin{subequations}
\begin{equation} \label{eqB1a}
\frac{\langle uw(z) \rangle_c}{u_{\tau}^2} = N_e \! \! \int^{1}_{z/\delta} \! \frac{d \zeta}{\zeta} I_{uw}(0,\zeta) .
\end{equation}
Here $I_{uw}$ is defined in the same manner as for $I_{u^n u^m}$ in Eq.~(\ref{eq11}). The condition $w=0$ at $z = 0$ implies $I_{uw} \rightarrow 0$ at $\zeta = z/h_e \rightarrow 0$ \cite{t76}. With use of some functions $\hat{I}_{uw}^{(l)}$, we expand $I_{uw}$ in a Maclaurin series as
\begin{equation} \label{eqB1b}
I_{uw} \! \! \left( \frac{r}{z}\zeta, \zeta \right) \! = \! \sum_{l=1}^{\infty} \hat{I}_{uw}^{(l)} \! \! \left( \frac{r}{z} \right) \frac{\zeta^l}{l!} .
\end{equation}
The integration of Eq.~(\ref{eqB1a}) is thus convergent at $z/\delta \rightarrow 0$,
\begin{equation} \label{eqB1c}
\frac{\langle uw(z) \rangle_c}{u_{\tau}^2} \rightarrow c_{uw} = N_e \! \! \int^{1}_{0} \! \frac{d \zeta}{\zeta} I_{uw}(0,\zeta) .
\end{equation}
Being analogous to Eq.~(\ref{eq1c}), this is a scaling law. We impose $c_{uw} = -1$ so as to be consistent with $\langle -uw \rangle = u_{\tau}^2$ \cite{t76}. However, if the limit $z/\delta \rightarrow 0$ is not taken, Eq.~(\ref{eqB1a}) is retained as follows \cite{pc82}:
\begin{align} \label{eqB1d}
\frac{\langle uw(z) \rangle_c}{u_{\tau}^2} &= c_{uw} - N_e \! \! \int^{z/\delta}_{0} \! \frac{d \zeta}{\zeta} I_{uw}(0,\zeta) \nonumber \\
                                           &= c_{uw} - \! \sum_{l=1}^{\infty} N_e \hat{I}_{uw}^{(l)}(0) \frac{( z/\delta )^l}{l! l} . 
\end{align}
\end{subequations}
The residual terms of $\varpropto (z/\delta)^l$ correspond to the aforementioned variation of $\langle -uw \rangle$. As for pipes and channels, we need $\hat{I}_{uw}^{(1)}(0) \ne 0$ and $\hat{I}_{uw}^{(2)}(0) = \hat{I}_{uw}^{(3)}(0) = ... = 0$. Since these are not necessarily equal to $0$ in boundary layers, the significance of each of the terms is likely to depend on the configuration of the flow.

Then, we consider $\langle u^n(x,z)u^m(x+r,z) \rangle_c /u_{\tau}^{n+m}$ in Eq. (\ref{eq3a}), which has been obtained from $I_{u^n u^m}$ in Eq.~(\ref{eq12}). Because of $u \ne 0$ at $z = 0$, the Maclaurin series of $I_{u^n u^m}$ has a term at $l = 0$ for $\hat{I}_{u^n u^m}^{(0)}$,
\begin{subequations} \label{eqB2}
\begin{equation} \label{eqB2a}
I_{u^n u^m} \! \! \left( \frac{r}{z}\zeta, \zeta \right) \! = \! \sum_{l=0}^{\infty} \hat{I}_{u^n u^m}^{(l)} \! \! \left( \frac{r}{z} \right) \frac{\zeta^l}{l!} .
\end{equation}
If the limit $z/\delta \rightarrow 0$ is not taken, Eq.~(\ref{eq12a}) is retained as
\begin{align} \label{eqB2b}
\! \frac{\langle u^n(x,z)u^m(x+r,z) \rangle_c}{u_{\tau}^{n+m}} \! & = C_{u^nu^m} \! \! \left( \frac{r}{z} \right) + D_{u^nu^m} \! \! \left( \frac{r}{z} \right) \! \ln \! \left( \frac{\delta}{z} \right) \! \nonumber \\
                                                                  & \quad  - \! \sum_{l=1}^{\infty} N_e \hat{I}_{u^n u^m}^{(l)} \! \! \left( \frac{r}{z} \right) \frac{( z/\delta )^l}{l! l}.
\end{align}
\end{subequations}
We have defined $C_{u^n u^m}$ in Eq.~(\ref{eq12b}) and $D_{u^n u^m} \varpropto \hat{I}_{u^n u^m}^{(0)}$ in Eq.~(\ref{eq12c}). Since Eq.~(\ref{eqB2b}) is not exactly logarithmic, it is concluded that the logarithmic law of Eq.~(\ref{eq3a}) holds only asymptotically in the limit $z/\delta \rightarrow 0$ of the constant-flux sublayer.

The same is true for $\langle \varepsilon^n(x,z) \varepsilon^m(x+r,z) \rangle_c / (u_{\tau}^3/z)^{n+m}$ in Eq.~(\ref{eq3b}). If the limit $z/\delta \rightarrow 0$ is not taken, we retain residual terms of $\varpropto (z/\delta)^l$. Such a term might explain the variation of $\langle \varepsilon (z) \rangle_c / (u_{\tau}^3/z)$ in Fig.~\ref{f1}(e), which is not consistent with Eq.~(\ref{eq3b}) or (\ref{eq1c}). We would need to study the constant-flux sublayer obtained at the smaller values of $z/\delta$ in a laboratory or in a field.

Finally, these results are reconsidered with our theory of Sec.~\ref{S2}. If the limit $z/\delta \rightarrow 0$ is not taken, $z/\delta$ is retained as a parameter in Eq.~(\ref{eq7c}),
\vspace{-1mm}
\begin{subequations} \label{eqB3}
\begin{equation} \label{eqB3a}
\frac{\partial}{\partial z} \ln \langle e^{is \alpha + it \beta} \rangle = - \frac{\varphi (s,t,r/z,z/\delta)}{z}  .
\end{equation}
Here $z/ \delta$ is independent of $s$, $t$, and $r/z$ but is not of $z$ because $\delta$ is a constant (Sec.~\ref{S1}). By utilizing the Maclaurin series of $\varphi^{(n,m,0,0)} (0,0,r/z,z/\delta)$, the corresponding extension of Eq.~(\ref{eq7d}) is written as
\begin{equation} \label{eqB3b}
\! \frac{\partial}{\partial z} \langle \alpha^n \beta^m \rangle_c  = - \sum_{l=0}^{\infty} \frac{ \varphi^{(n,m,0,l)} (0,0,r/z,0)}{i^{n+m} z} \frac{(z/\delta)^l}{l!}. \!
\end{equation}
\end{subequations}
The integration of Eq.~(\ref{eqB3b}) leads to $\langle \alpha^n \beta^m \rangle_c$. However, there are again residual terms of $\varpropto (z/\delta)^l$. For the case of $\alpha = [U(z)+u(x,z)]/u_{\tau}$ and $\beta = [U(z)+u(x+r,z)]/u_{\tau}$ in Eq.~(\ref{eq4}), we expect that $\varphi^{(n,m,0,l)} (0,0,r/z,0)/i^{n+m}$ in Eq.~(\ref{eqB3b}) is identical to $N_e \hat{I}_{u^n u^m}^{(l)}(r/z)$ in Eq.~(\ref{eqB2b}). A similar identity would hold even if other quantities are used for the variables $\alpha$ and $\beta$.

\end{document}